\documentclass[a4paper,12pt]{article}
\pdfoutput=1
\usepackage{amssymb,amsmath,bm}
\usepackage{color}
\usepackage{slashed}
\usepackage[latin1]{inputenc}
\usepackage{amsfonts}
\usepackage{lscape}
\usepackage{amsthm}
\usepackage{booktabs}
\usepackage{array}
\usepackage{rotating}
\usepackage{float}
\usepackage{multirow}
\usepackage{graphicx,rotating,amsmath,multirow,xcolor,colortbl,xcolor}
\usepackage{hyperref,pdflscape,slashed}
\definecolor{rosso}{cmyk}{0,1,1,0.4}
\definecolor{rossos}{cmyk}{0,1,1,0.55}
\definecolor{rossoc}{cmyk}{0,1,1,0.2}
\definecolor{blu}{cmyk}{1,1,0,0.3}
\definecolor{blus}{cmyk}{1,1,0,0.6}
\definecolor{bluc}{cmyk}{1,1,0,0.1}
\definecolor{verde}{cmyk}{0.92,0,0.59,0.25}
\definecolor{verdec}{cmyk}{0.92,0,0.59,0.15}
\definecolor{verdes}{cmyk}{0.92,0,0.59,0.4}
\definecolor{Gray}{gray}{0.95}

\font\tenrsfs=rsfs10 at 12pt
\font\sevenrsfs=rsfs7
\font\fiversfs=rsfs5
\newfam\rsfsfam
\textfont\rsfsfam=\tenrsfs
\scriptfont\rsfsfam=\sevenrsfs
\scriptscriptfont\rsfsfam=\fiversfs
\def\mathscr#1{{\fam\rsfsfam\relax#1}}

\usepackage[small,bf]{caption}
\def\hhref#1{\href{http://arxiv.org/abs/#1}{#1}} 
\hypersetup{colorlinks,bookmarksopen,bookmarksnumbered,
linkcolor=blus,pdfstartview=FitH,urlcolor=rossos,citecolor=verde}

\newcommand{\gev}{{\rm GeV}}

\newcommand{\lsim}{\stackrel{<}{_\sim}}

\newcommand{\be}{\begin{equation}}
\newcommand{\ee}{\end{equation}}
\newcommand{\bea}{\begin{eqnarray}}
\newcommand{\eea}{\end{eqnarray}}
\newcommand{\beq}{\begin{equation}}
\newcommand{\eeq}{\end{equation}}
\newcommand{\beqa}{\begin{eqnarray}}
\newcommand{\eeqa}{\end{eqnarray}}

\def\eq#1{eq.~(\ref{#1})}
\def\fig#1{fig.~\ref{#1}}
\def\sect#1{sect.~\ref{#1}}
\def\tab#1{table~\ref{#1}}

\def\Eq#1{Eq.~(\ref{#1})}

\begin{document}

CERN-TH-2016-106

\vspace{1truecm}

\begin{center}
\boldmath

{\textbf{\LARGE Hunting for Dark Particles\\ 
\vspace{0.2 truecm}
with Gravitational Waves
}}
\unboldmath

\bigskip

\vspace{0.2 truecm}

{\bf Gian F. Giudice, Matthew McCullough, and Alfredo Urbano}
 \\[5mm]

{\it CERN, Theoretical Physics Department, Geneva, Switzerland}\\[2mm]

\end{center}

\vspace{1cm}

\begin{quote}
\noindent
The LIGO observation of gravitational waves from a binary black hole merger has begun a new era in fundamental physics.  If new dark sector particles, be they bosons or fermions, can coalesce into exotic compact objects (ECOs) of astronomical size, then the first evidence for such objects, and their underlying microphysical description, may arise in gravitational wave observations.  In this work we study how the macroscopic properties of ECOs are related to their microscopic properties, such as dark particle mass and couplings.  We then demonstrate the smoking gun exotic signatures that would provide observational evidence for ECOs, and hence new particles, in terrestrial gravitational wave observatories.  Finally, we discuss how gravitational waves can test a core concept in general relativity: Hawking's area theorem.
\end{quote}

\thispagestyle{empty}
\vfill

\vspace{-0.5cm}

\newpage

\tableofcontents
\newpage

\section{Introduction}

The LIGO detection of GW150914~\cite{Abbott:2016blz} -- the first observed event of gravitational wave (GW) emission from binary black-hole merging -- has ushered in a new era in the exploration of our universe. Not only does this result open new prospects for the study of the astrophysical properties of objects in extreme density conditions, but also allows for new ways of testing fundamental physics. Indeed, soon after the LIGO announcement, it was suggested that GW observations can be used to probe violations of the equivalence principle and Shapiro delay~\cite{Kahya:2016prx,Wu:2016igi,Yunes:2016jcc,Liu:2016edq}, modifications of gravity~\cite{Yunes:2016jcc,Konoplya:2016pmh,Moffat:2016gkd,Vainio:2016qas}, the presence of event horizons~\cite{Cardoso:2016rao}, the quantum structure of black holes~\cite{Giddings:2016tla}, the hypothesis that dark matter is in the form of primordial black holes~\cite{Sasaki:2016jop,Bird:2016dcv,Clesse:2016vqa}, the propagation of gravity waves~\cite{Collett:2016dey,Bicudo:2016pps,Blas:2016qmn,Calabrese:2016bnu,Garcia-Bellido:2016zmj,Schreck:2016qiz,Yunes:2016jcc,Arzano:2016twc,Branchina:2016gad}, phase transitions in the early universe~\cite{Dev:2016feu,Jaeckel:2016jlh}, neutrino properties from associated neutrino emission~\cite{Langaeble:2016han,Lehner:2016lxy}, and axion clouds around black holes~\cite{Arvanitaki:2016qwi}.

In this paper we want to investigate another way in which GW detection can contribute to fundamental science: the search for new physics in the form of exotic compact objects (ECO).  Dark matter is a primary motivation to expect that new forms of stable matter must exist and be abundant in our universe. Moreover, quite independently of dark matter considerations, many new-physics theories predict the existence of stable or long-lived light particles with feeble interactions with ordinary matter. Because of their elusive nature, here we will generally refer to them as dark particles. It is conceivable that dark particles could coalesce into macroscopic objects. Although at present we are not able to predict the mass distribution of such exotic objects, it is possible that they could reach astronomical sizes. In this case GW detection from the merging of two ECOs or the collapse of an ECO into a stellar black hole could be used as a new and formidable way to probe the existence of dark particles and test their properties. Actually, in some cases, this could be the only way in which dark particles are discovered, because their extremely feeble interactions make them completely invisible to collider searches or other particle physics experiments. 

This paper is organised as follows. In \sect{sec:exo} we describe scenarios involving new particle physics beyond the Standard Model which can give rise to ECOs, focussing on properties relevant to GW signatures. In \sect{signatures} we explain how ECOs could be identified through GW detection because of unexpected features either in the gravitational waveforms or in the distribution of observed events. In \sect{sec:area} we explore how GW observations allow for tests of Hawking's area theorem~\cite{Hawking:1971tu}, and thus of some of the fundamental principles underlying general relativity.  In \sect{sec:conc} we summarise our results.

Throughout the paper, we will use experimental parameters corresponding to Advanced LIGO at design sensitivity which, for simplicity, will be referred to as LIGO. Furthermore,
we will work in natural units in which Planck's constant $\hbar$, the speed of light $c$, and Newton's constant $G_N$ are set to one. Occasionally, we will reintroduce these constants to make some equations more transparent.

\section{Features of Exotic Compact Objects}
\label{sec:exo}

ECOs are predicted in a variety of new-physics theories and are motivated by very different physics considerations. Here we will systematically analyse the various possibilities, trying to classify the cases according to their common astrophysical features, rather than their underlying physics motivation or theory implementation. 

From the astrophysical point of view, ECOs present some universal properties, behaving as collisionless bodies in the same way as stars.  If they do not emit significant electromagnetic radiation, they escape most astronomical searches.  The strongest observational constraints arise from  microlensing events, where light from a distant source is distorted whenever a compact object passes near the line of sight.   This creates unresolved multiple images observable as an amplification. These searches have been performed for the broad class of Massive Compact Halo Objects (MACHOs) and apply to the ECOs considered here.  It has been determined that compact objects in the mass range $0.6\times10^{-7} \ M_{\odot} < M <15\ M_{\odot}$ cannot be the primary component of the Milky Way halo and, for $M <10\ M_{\odot}$ can comprise at most 20\% of the dark matter mass~\cite{Alcock:2000ph,Tisserand:2006zx}.  This is a very powerful result from the perspective of dark matter, but it shows that ECOs may be numerous and may well account for as much matter as is observed in the form of baryons. 

\subsection{Conventional Compact Objects}
\label{sec:conv}
Before discussing ECOs and the way to distinguish them from known astrophysical compact objects, such as black holes (BH) and neutron stars (NS), it is useful to review briefly the most relevant features of the ``background" to our ``new-physics signal" (to use particle physics terminology).

NS are the supernova remnant of massive stars in which electrons disappear through inverse $\beta$ decay. The gravitational attraction is balanced by the Fermi pressure of degenerate neutrons. The maximum NS mass depends on unknown parameters in the equation of state (EoS). However, robust upper limits on NS masses can be derived, independently of the EoS, from basic dynamical assumptions such as the Oppenheimer-Volkoff equation and the causality condition that the speed of sound is less than the speed of light. In this way, one obtains the following upper limits on NS masses:
\bea
M < 3.2~M_\odot ~~~~~&&\hbox{Rhoades-Ruffini limit~\cite{Rhoades:1974fn}}
\nonumber \\
M < 3.6~M_\odot ~~~~~&&\hbox{Nauenberg-Chapline limit~\cite{Nauenberg}}.  
\eea 
These limits are derived for non-spinning NS, and the rotation effect increases the maximum mass by at most 20\%, leading to $M < 4.3~M_\odot$. However, these are theoretical upper bounds. The most massive NS ever observed is PSR J0348+0432 with $M = 2.01\pm0.04~M_\odot$~\cite{Antoniadis:2013pzd}. Numerical NS models with realistic EoS can hardly exceed values about $2 M_\odot$. These upper bounds on NS masses allow one to distinguish NS from BH in GW events: for conventional forms of matter, any compact object of mass larger than about 3--4 solar masses must be a BH. 

Another important parameter that differentiates NS from BH is compactness, defined as the ratio between an object's mass and radius. For 
non-rotating BH, compactness is given by
\beq
C\equiv \frac{M}{R}=\frac{1}{2}~~~~~\hbox{(for BH)}.
\label{compBH}
\eeq

For NS, an absolute upper bound on compactness, independent of the EoS, is given by~\cite{Buchdahl:1959zz, weinberg}
\beq
C\equiv \frac{M}{R}<\frac49 ,~~~~~~~~~~\hbox{(for NS)},
\eeq
which can be saturated for an incompressible star with uniform density. A more stringent upper bound, although dependent on EoS assumptions,  has been obtained by Lindblom~\cite{lindblom}: $M/R<0.35$. However, with realistic assumptions on the EoS, one finds that the typical range of NS compactness is $0.13\lsim M/R \lsim 0.23$~\cite{shapiro,Lattimer:2006xb}.

We see that, as in the case of the mass distribution, also for compactness there is a gap between the NS and BH range. This is another important element to distinguish NS from BH in GW signals. Indeed, as the stars in the binary approach each other during the inspiral phase at distances comparable to $R$, the EoS starts to influence the GW emission. In particular, the tidal deformations in NS binaries modify the orbital evolution and leave a signature in the transition between inspiral and merger phases~\cite{Andersson:2009yt}. 

\subsection{Boson Stars}
\label{sec:exobs}

Boson stars (for reviews, see~\cite{Jetzer:1991jr,Liddle:1993ha,Schunck:2003kk,Liebling:2012fv}) are often considered as a template for exotic stars because of the relative simplicity of the equations that govern their dynamics and the richness in the physical phenomena that they can exhibit. After the discovery of the Higgs boson and the ascent of the inflaton as a crucial ingredient of cosmological theories, the reasons for new scalar fields have been strengthened significantly. Boson stars may not be just an academic exercise, but part of our universe.
 
There is no lack of theories in which new bosonic degrees of freedom find a natural setting. Examples are the axion as a solution of the strong CP problem, flat directions in supersymmetric models, moduli in string theory, a variety of pseudo-Goldstone bosons, new light vector fields, or dark matter candidates in different contexts.

The gravitational pull in a boson star is neither supported by thermal pressure (as in luminous stars) nor Fermi degeneracy (as in white dwarfs or NS). Its stability is provided by the quantum property that particles cannot be localised beneath distances of the order of their Compton wavelength. Indeed, the Heisenberg principle dictates that the localisation length $R$ of a particle with mass $m_B$ and momentum $p=m_Bc$ must satisfy $R>\hbar /(m_Bc)$. When this limit is saturated for the smallest Kepler orbit before gravitational collapse (given by $3R_S$, where $R_S=2G_N M/c^2$ is the Schwarzschild radius), we find that the maximum mass of a boson star is about $M_{\rm max} \approx M_P^2/m_B$, where $M_P$ is the Planck mass.
This shows that a boson star is a macroscopic quantum object that effectively behaves like a Bose-Einstein condensate of astronomical size.

The above estimate of the largest mass of a boson star is in good agreement with the solution of the general relativistic equations, which give~\cite{Kaup:1968zz}
\beq
M_{\rm max} =0.633 ~\frac{M_P^2}{m_B} \approx \left( \frac{10^{-10}~{\rm eV}}{m_B}\right) M_\odot
~~~~~\hbox{(for free-field boson stars)}.
\label{bosstar}
\eeq
This shows that GW signals from solar mass-scale objects are observable by LIGO for extremely light boson fields. 
The correct range of masses can be obtained in the case of the QCD axion, for an axion decay constant in the range of GUT or string scales, and in a variety of models of axion-like particles involving new physics. Recently it has also been shown that boson stars may also form out of massive vector fields \cite{Brito:2015pxa}, although we do not consider this case here.

 The result in \eq{bosstar} changes completely as soon as new forces come into the game. For instance, let us include a quartic coupling in the potential $V$ for the complex boson field $\phi$,
 \beq\label{eq:BosonPotential}
 V(\phi )  =m_B^2 |\phi |^2 +\frac{\lambda}{2} |\phi |^4 .
 \eeq
This gives a repulsive self-interaction that acts as an extra source of pressure against gravitational collapse. The maximum mass of this self-interacting boson star is \cite{colpi}
\beq
M_{\rm max} =0.06 \sqrt{\lambda} ~\frac{M_P^3}{m_B^2} \approx \sqrt{\lambda} \left( \frac{100~{\rm MeV}}{m_B}\right)^2 10~M_\odot
~~~~~\hbox{(for self-interacting boson stars)}.
\label{bosstar2}
\eeq
The value of $M_{\rm max}$ is now parametrically equal to the Chandrasekhar mass corresponding to a constituent fermion of mass $m_B/\lambda^{1/4}$. The result  in \eq{bosstar2} suggests that GW relevant for LIGO may be obtained, in the self-interacting case, for values of $m_B$ in the MeV to GeV range.

An alternative mechanism to stabilise a boson star is through non-topological solitons~\cite{rosen,Friedberg:1976me,Coleman:1985ki,Lee:1991ax}. Unlike topological solitons (such as magnetic monopoles~\cite{'tHooft:1974qc,Polyakov:1974ek}) which have non-trivial asymptotic configurations at spatial infinity different from the ordinary vacuum state, non-topological solitons are localised solutions of the equations of motion with trivial asymptotic behaviour. The existence of such solutions for a scalar field is possible in the presence of a conserved charge $Q$. In practice, a lump of bosons with large $Q$ can be stable if its energy per unit charge decreases at larger $Q$, so that the bound state is energetically favoured with respect to isolated charges. Such lumps, called $Q$-balls, are stable and localised even in the absence of gravity.

Once gravity is included, one obtains a maximum mass for the corresponding star. The result depends on the form of the potential that is used to obtain the non-topological soliton. For instance for a scalar field with mass $m_B$ and condensate $\phi_0$, one finds $M_{\rm max}\approx M_P^4/(m_B\phi_0^2)$ for the potential used in ref.~\cite{Lee:1986ts,Friedberg:1986tp,Friedberg:1986tq} and $M_{\rm max}\approx M_P^3/\phi_0^2$ for $Q$-stars~\cite{Lynn:1988rb}. Thus, it is possible for $Q$-balls to reach tens of solar masses and high densities. The existence of macroscopic $Q$-balls is common in a variety of new-physics models with scalar fields that carry a conserved quantum number and develop a condensate. In particular, this can happen in supersymmetric models, in which the scalar field corresponds to a flat direction and the conserved charge could be lepton number, baryon number, or a combination of the two~\cite{Enqvist:2003gh,Dine:2003ax}.   

In the case of the quartic self-interaction, it is found~\cite{AmaroSeoane:2010qx} that the maximum compactness grows with the coupling constant $\lambda$ and saturates at strong coupling at the value 
\beq
(M/R)_{\rm max} =0.16
~~~~~\hbox{(for self-interacting boson stars)}.
\label{bosstar3}
\eeq

In the non-interacting case, the general relativistic solutions give a continuous profile for the scalar field and boson stars do not have a well defined surface. Nonetheless, one can define an effective radius of the star as the distance that envelops an appropriate fraction of the total mass (usually taken to be 99\%). With this definition, it is possible to compute the effective compactness of a boson star and one finds
\beq
(M/R)_{\rm max} =0.08
~~~~~\hbox{(for free-field boson stars)}.
\label{bosstar4}
\eeq

\Eq{bosstar3} places the maximum compactness of self-interacting boson stars in the same ballpark of NS. 

\subsection{Fermion Stars}
\label{sec:exofs}

Even without invoking new physics, unconventional states of matter could exist and form compact objects with GW signatures different from those of BH or NS. At very high density, the boundaries between nucleons may dissolve and nuclear matter transform into quark matter. The fate of this quark matter in compact stars is not fully understood. It may be in the form of a colour superconducting state~\cite{Alford:1997zt,Rapp:1999qa,Alford:2001dt}, in which quarks are arranged in Cooper pairs that necessarily carry colour. Or it may be in the form of strange matter~\cite{Itoh:1970uw,Bodmer:1971we,Witten:1984rs,Alcock:1986hz,Madsen:1998uh,Weber:2004kj}, in which the addition of  strange quarks to a medium made of up and down quarks (together with a component of electrons necessary to neutralise electric charge) lowers the energy by increasing the degree of degeneracy, creating a system more stable than nuclear matter. 

If any form of quark matter really exists, the structure of NS could be quite different than what is ordinarily believed. NS would be largely composed of colour superconducting quark matter or strange matter, with at most a thin external layer of degenerate neutrons. These differences in the EoS modify the characteristic frequency of tidal disruption, leading to observable effects in the shape of the GW emission at the end of inspiral. This has been confirmed by studies of GW signals from binaries with strange stars~\cite{Limousin:2004vc,Lattimer:2006xb,GondekRosinska:2008nf,Bauswein:2009im}.

Exotic kinds of stars emerge in the context of
many new physics theories that predict the existence of a ``mirror world"~\cite{Berezhiani:2003xm} -- a sort of duplication of the SM with new particles and forces.
Perhaps the best-motivated scenario of this class arises in Twin Higgs models~\cite{Chacko:2005pe}.  In these models the little hierarchy problem is solved by realising the Higgs boson as a pseudo-Goldstone boson of an approximate $\text{SU}(4)$ global symmetry.  This global symmetry is protected at the quantum level by a ${Z}_2$ exchange symmetry between SM fields and fields of the Twin sector, leading to a quadratic Higgs potential that accidentally respects the $\text{SU}(4)$ global symmetry at all orders in perturbation theory.  Since this exchange symmetry must be respected to a reasonable level in order to preserve the success of the model, the Twin sector may resemble the SM to quite a high degree.  In fact, in the extreme scenario dubbed the `Identical Twin'~\cite{Chacko:2005pe}, the field content is identical to the SM.  This possibility is ruled out by cosmological constraints on additional $\text{U}(1)$ gauge fields. However, if the Twin photon is given a small mass, then the required interactions between twin nucleons, twin electrons, and a light twin photon may allow for sufficient dissipative cooling to allow twin nucleons to collapse and form a twin neutron star.

While it has not been robustly demonstrated that a Twin Higgs sector may lead to the formation of compact objects and remain consistent with all cosmological bounds, phenomenologically similar models have been extensively studied.  The recently proposed DDDM framework \cite{Fan:2013tia,Fan:2013yva} and also other models of dissipative hidden sectors \cite{Foot:2014uba} exhibit the required properties to give rise to ECOs.  In fact, the possibility of hidden sector compact objects was raised in the context of these studies \cite{Fan:2013yva,Foot:2014uba}.  One of the toy models proposed in \cite{Fan:2013tia,Fan:2013yva} involves a heavy fermion which behaves analogously to the proton, a light fermion analogous to the electron, and a dark $\text{U}(1)$ gauge group analogous to electromagnetism.  With these ingredients it is not unrealistic to speculate about the possibility of the formation of ECOs.

Quite generally, we can view the mirror world as a duplication of the SM in which the parameters relevant to atomic and nuclear physics (such as electron, neutron, and proton masses, the QCD scale parameter, the QED fine-structure constant, or neutrino couplings) can take values different than in our world. This opens up the construction of a variety of stellar objects with characteristics quite distinct from conventional stars. In particular, it is possible to imagine exotic populations of supermassive stars, since the mass $m_F$ of the new fermionic particle entering the Chandrasekhar limit ($M\lsim M_P^3/m_F^2$) is now a free parameter, which could be much smaller than the ordinary value in nuclear physics.

Note that such mirror stars would be quite difficult to observe directly. Their electro\-magnetic-like or neutrino-like emissions would not be in the conventional form of photons and neutrinos, but in their mirror counterparts, which are most likely undetectable. In this respect, mirror stars behave like ECOs and GW emissions from their merging in binary systems may be the most promising way to hunt for them.

There are a variety of interesting additional cosmological signatures and, as long as these mirror ECOs only comprise some fraction of the total observed dark matter, they can be consistent with current bounds.  Perhaps the strongest bounds come from hidden sector acoustic oscillations \cite{Cyr-Racine:2013fsa}.  Constraints coming from the CMB and large scale structure observations require that for typical parameters dissipative hidden sectors should not comprise more than $\sim 5\%$ of the total dark matter \cite{Cyr-Racine:2013fsa}.

\subsection{Dark Matter Stars}
\label{sec:exodm}

Perhaps the strongest motivation for exotic new particles is due to the overwhelming evidence for dark matter.  In many well motivated new-physics scenarios, such as the axion solution of the strong-CP problem or supersymmetric theories addressing the hierarchy problem, compelling dark matter candidates arise.  These dark matter candidates are particularly attractive as they are cold and collisionless, thus already demonstrating the basic behaviour expected of particle dark matter.  

However, there is emerging evidence that dark matter may not be collisionless, but instead self-interacting. Indications in this direction come from numerical simulations of non-interacting dark matter, which predict excessively cuspy profiles of dwarf galactic halo cores (known as the ``core-cusp problem"~\cite{deBlok:2009sp}), a too-large number of satellite galaxies (known as the ``missing satellite problem"~\cite{Klypin:1999uc,Moore:1999nt}), and dwarf galaxies too massive to not have visible stars (known as the ``too big to fail problem"~\cite{BoylanKolchin:2011de}), in contradiction with observations. Moreover, studies of elliptical galaxies falling into the Abell 3827 galaxy cluster, with galactic dark matter subhaloes tracked by gravitational lenses, seem to indicate a non-vanishing dark matter self-interaction cross section~\cite{Massey:2015dkw}. 

In order to solve these problems in dwarf galactic haloes, one needs a cross-section per unit dark-matter mass
in the range
\begin{equation}\label{eq:DwarfCrossSection}
\sigma/m_{\rm DM} \simeq [0.1-10]~~{\rm cm}^2/{\rm g}~~~~~~\hbox{Dwarf haloes~\cite{Vogelsberger:2012ku,Rocha:2012jg,Zavala:2012us,Peter:2012jh}}~.
\end{equation}
The most stringent constraint on dark matter self-interaction comes from  observations in Milky Way-like elliptical galaxies, since 
too large cross sections would lead to spherical rather then elliptical haloes
\begin{equation}\label{eq:GalaxyCrossSection}
\sigma/m_{\rm DM} \lesssim  0.1-1~~{\rm cm}^2/{\rm g}~~~~~~\hbox{Milky Way-like haloes~\cite{Vogelsberger:2012ku,Rocha:2012jg,Peter:2012jh}}~.
\end{equation}
Thus, a momentum-independent self-interaction cross-section (generated for instance from a contact quartic interaction in the case of scalar dark matter) in the ballpark $\sigma/m_{\rm DM} \simeq [0.1-1]$ cm$^2$/g seems to satisfy  
all the astrophysical constraints. 

The above considerations, together with the emergence of new ideas in theoretical model building, has led to increasing interest in more complex dark matter candidates, which in many cases exhibit non-trivial self interactions.
In this respect, an interesting possibility is the case in which dark matter self-interactions are mediated by a light force carrier. 
While a contact interaction implies a constant self-interaction cross section, the presence of a light mediator 
introduces a non-trivial velocity dependence. As a rule of thumb, in analogy with Rutherford scattering, 
$\sigma/m_{\rm DM}$ is enhanced at low velocity, and suppressed at high 
velocity. This has important consequences in light of the astrophysical bounds discussed before since 
galactic haloes of different size are characterised by different dispersion velocities, usually inferred from the measurement of rotation curves.
For instance, a dwarf halo has a typical dispersion velocity  of $v_0 \simeq 10$ km/s,
whereas for a Milky Way-like halo we have $v_0 \simeq 200$ km/s.
It is therefore possible that self-interactions are manifest in smaller halos,
while dark matter appears to be collisionless on larger scales. This effect will be important for our discussion of LIGO sensitivity to fermion stars in \sect{sec:Frequency}.

Furthermore, if the mediators are sufficiently light they may enable self-gravitating dark matter clouds to collapse and it has been demonstrated in \cite{Kouvaris:2015rea} that under certain circumstances, for asymmetric fermionic dark matter, the collapsed cloud may form a compact dark star supported by fermion degeneracy pressure. The characteristics of these dark-matter motivated ECOs are similar to those described in sects.~\ref{sec:exobs}, for bosonic dark matter, and \ref{sec:exofs}, for fermonic dark matter. In particular, ECOs with masses in the range interesting for GW observations can be obtained for sub-GeV dark matter particles, with the exact value depending on the self-interaction parameters. A detailed study of the maximum mass of stars made of asymmetric fermionic dark matter can be found in~\cite{Kouvaris:2015rea}. 

The microlensing limits on the density of galactic compact objects discussed at the beginning of this section can be easily satisfied by assuming that only a fraction of dark matter forms ECOs, while the majority remains as free particles, just like the case of baryonic matter, for which stars and dust can coexist in the universe. Computing the fraction of dark matter in the form of ECOs would require more knowledge of the processes for dark-matter star formation than presently available. 

\subsection{Multi-Component Stars}

While so far we have considered only objects constituted by a single new particle, a variety of ECOs could be formed by a combination of several new particles or mixtures of new and ordinary particles. These cases are particularly interesting in the context of dark matter models. Boson stars could exist in the core of ordinary or exotic fermion stars, or vice versa. The various components could couple only gravitationally or also through new interactions~\cite{Henriques:1989ar,Henriques:1989ez,Jetzer:1990xa,Henriques:1990xg,deSousa:2007di,Brito:2015yfh,Brito:2015yga}. A theoretically intriguing situation is the one of a non-topological soliton generated by a scalar field that gives mass to a fermion~\cite{Lee:1986tr,Chiu:1990zz}. As a result of the non-trivial (and position dependent) configuration of the scalar field, the fermion could be massless inside the boson star, but massive outside. 

An interesting interplay between dark and ordinary matter occurs in the so-called Dark Stars (for a review, see~\cite{Freese:2015mta}). Dark Stars are stellar objects made by ordinary atomic matter and fuelled by the heat produced by dark matter annihilation in their cores. This process can successfully support the stability of the star for dark matter annihilation cross section consistent with what predicted from thermal relic abundance calculations. It is conceivable that Dark Stars can accrete and become very massive, much more massive than ordinary stars. However, Dark Stars are not very compact and it is unlikely for them to generate detectable GW signals. On the other hand, they could be observed via more conventional astronomical techniques~\cite{Freese:2015mta}.  Similarly, an interesting interplay between annihilating dark matter and black holes is possible \cite{Bertone:2005xz,Lacki:2010zf}, where dark matter annihilations in the high density environment of a black hole could lead to visible signatures of dark matter.  However, once again this is not linked specifically to GW signatures.

\subsection{Dark Energy Stars}
\label{sec:gravastar}

Dark energy provides a new substance to form a completely different kind of ECOs. One can envisage celestial bodies made of macroscopic bubbles of vacuum energy, whose negative pressure is compensated by an external shell of fluid with ordinary attractive gravitational pull. The simplest example of this class of ECOs is the {\it gravastar} (from GRAvitational VAcuum STAR, $g^*$ hereafter)~\cite{Mazur:2001fv}. 

A $g^*$ is described by an internal spherical core of radius $r_c$ with constant energy density $\rho_0$ and negative pressure $p=-\rho_0$, and an external shell of thickness $R-r_c$ made of an ultrarelativistic fluid characterised by the equation of state $p=\rho$. The profile of the energy density $\rho$ in the shell region is model-dependent. However, it can be determined by the Oppenheimer-Volkoff equation in terms of $\rho_0$, once we impose continuity conditions for $\rho$ and its derivative at $r=r_c$ and the boundary condition $\rho(R) =0$. So the simplest model of a $g^*$ is described by a total of three parameters, and has a discontinuity in $p$ at the interface, where $r=r_c$. The pressure discontinuity can be resolved by replacing the zero-thickness interface with a thin crust with anisotropic pressure~\cite{Cattoen:2005he}.

Gravastar stability under thermodynamic conditions, spherically symmetric perturbations, or rotational effects  has been studied by several authors~\cite{Visser:2003ge,Carter:2005pi,Lobo:2005uf,Chirenti:2007mk,Chirenti:2008pf}, and it has been demonstrated that in the ultra-compact limit they can become gravitationally unstable \cite{Cardoso:2014sna}.

It remains to be proven that the merging of two $g^*$ leads to a new stable $g^*$, or even that they can form in the first place, however for the purposes of studying the GW signatures of $g^*$ mergers we will assume both to be possible here.

\section{Signatures of Exotic Compact Objects}
\label{signatures}

\subsection{LIGO Sensitivity on ECO parameters}\label{sec:Frequency}

Let us consider two point masses $M_1$ and $M_2$ orbiting each other under the force of gravity.  
This is equivalent to a single body with reduced 
mass $\mu \equiv M_1 M_2/M_{\rm tot}$ (where $M_{\rm tot} = M_1 + M_2$) moving in an external gravitational potential.  
The equivalent body moves in an elliptic orbit with major semi-axis $l$, describing the separation of the two bodies.

The orbital period $P$ is related to $M_{\rm tot} $ and $l$ by Kepler's third law 
\begin{equation}\label{eq:Freqbasta}
P^2=\frac{4\pi^2 l^3}{M_{\rm tot}} \, .
\end{equation}

The frequency $f$ of GW emission is twice the orbital frequency $\nu =1/P$, and thus is given by
\begin{equation}\label{eq:Freq}
f = \sqrt{\frac{M_{\rm tot} }{\pi^2 l^3}}~.
\end{equation}

For the system of two BHs the innermost stable circular orbit (ISCO) \cite{Ajith:2009bn} is defined by
\begin{equation}\label{eq:FreqBH}
R^{\rm ISCO}_{BH} \equiv 6M_{\rm tot} ~.
\end{equation}
The ISCO determines the end of the inspiral phase and the beginning of the merger phase, thus this radius characterises the typical frequency $f^{\rm ISCO}_{BH}$ expected in a binary merger. For a BH-BH merger, we find
\beq\label{eq:Freq1}
f^{\rm ISCO}_{BH} =\frac{1}{6^{3/2 }\pi M_{\rm tot}}  ~~~~~~{\rm (for~BH)} \, .
\eeq
The maximum frequency at the end of inspiral, obtained from
numerical simulations of the waveform of BH mergers, is given by  $f=f^{\rm ISCO}_{BH}(1+\Delta )$, where the correction term $\Delta$ is computed in post-Newtonian approximation and is a function of the mass ratio $M_1/M_2$ and of a single combination of the two BH spins. Its analytical expression can be found in \cite{Ajith:2009bn}. The correction term $\Delta$ vanishes in the limit of negligible BH spin and large mass ratio. For non-rotating BHs, it reaches a maximum value $\Delta=1.1$ for $M_1/M_2=3.56$, while $\Delta=0.7$ for equal BH masses. For spinning BHs, in the parameter range for which the expansion given in \cite{Ajith:2009bn} is valid, the maximum value reached is $\Delta=3.1$.  Thus, for discussion purposes, it is adequate to take the expression in \eq{eq:Freq1} as the frequency determining the end of inspiral. The correction term $\Delta$ is fully included in our numerical analysis.

By analogy with \eq{eq:FreqBH}, for ECOs we may define a typical GW frequency determined by the ISCO radius $R^{\rm ISCO}_{ECO} \equiv 3 \,M_{\rm tot}/C $, where $C \equiv M/R$ is the compactness of the ECO, assumed to be the same for the two merging bodies.  This ansatz is corroborated by numerical analyses of NS orbits \cite{Lai:1996sv}.

Consequently, we may define the typical frequency $f^{\rm ISCO}_{ECO}$ in the merging of two ECOs, using \eq{eq:Freq} for $l = 3 M_{\rm tot}/C$:
\begin{equation}\label{eq:Freq2}
f^{\rm ISCO}_{ECO} = \frac{C^{3/2}}{ 3^{3/2} \pi M_{\rm tot}} ~~~~~~{\rm (for~ECO)} \, .
\end{equation}
Note that \eq{eq:Freq2} becomes equal to \eq{eq:FreqBH}, when we approach the Schwarzschild BH compactness $C_{BH} = 1/2$.

\begin{figure}[!htb!]
\centering
  \includegraphics[width=.6\linewidth]{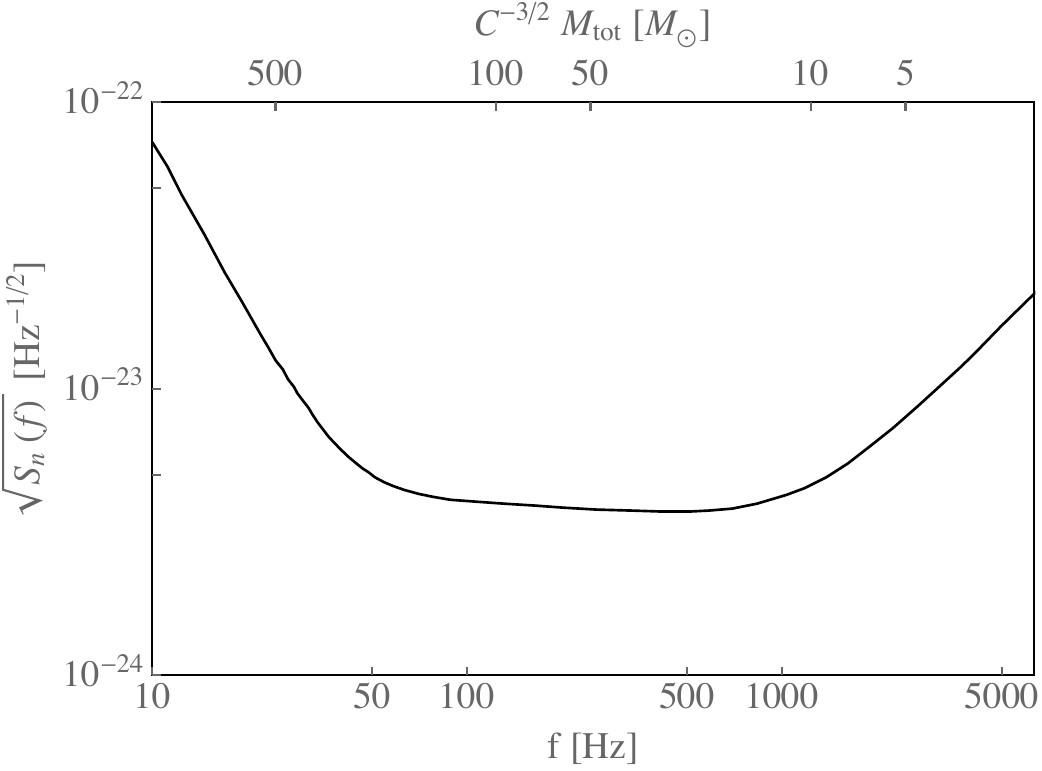}
\caption{\em 
LIGO noise power spectral density, taken from \cite{LIGOREF}.  For a given characteristic GW frequency for a binary merger, on the upper axes we show the corresponding combination of ECO compactness $C = M/R$ and total mass $M_{\rm tot}$.}
\label{fig:Noise}
\end{figure}

To observe the merger and ringdown phase 
it is crucial that the characteristic frequency $f^{\rm ISCO}$ falls within the LIGO sensitivity range.  LIGO sensitivity can be described in terms of the so-called noise power spectral density~\cite{Ajith:2011ec}.  We plot this function in \fig{fig:Noise}. The most sensitive interval, where the signal-to-noise is maximised, is $f = [50-1000]$ Hz.
On the upper axes of \fig{fig:Noise} we also show the sensitivity in terms of $C$ and $M_{\rm tot}$ by employing \eq{eq:Freq2}.  Thus we see that the signal-to-noise ratio is maximised for objects in the 
range 
\begin{equation}\label{eq:Range}
\left( \frac{C}{0.2}\right)^{3/2} 1.1\,M_\odot \lesssim M_{\rm tot} \lesssim \left( \frac{C}{0.5}\right)^{3/2} 88\,M_\odot~.
\end{equation}

For a GW signal with strain $h(t)$ the expectation value of the signal-to-noise ratio is
\begin{equation}
\rho^2 = 4\int_0^{\infty} \frac{|\tilde{h}(f)|^2}{S_n(f)}df~,
\label{intrho}
\end{equation}
where $\tilde{h}(f)$ is the Fourier transform of the signal. As we are interested in the region of best sensitivity, in \eq{intrho} we have taken the case of optimal orientation (face-on, overhead source). Averaging over sky position, inclination, and polarisation, one finds that $\langle \rho^2 \rangle$ is equal to $4/25$ times the expression in \eq{intrho}.
A criterion for detectability is the requirement $\rho \geqslant 8$~\cite{Dominik:2014yma}.
Assuming that most of the signal-to-noise ratio is accumulated during the inspiral phase, 
we can use the quadrupole approximation truncated at the Newtonian order~\cite{Khan:2015jqa}
\begin{equation}\label{eq:Strain}
\tilde{h}(f) \approx \frac{\sqrt{5/24}}{\pi^{2/3} D_L}M_{c}^{5/6}f^{-7/6}~,
\end{equation}
where 
\be
M_c = \frac{(M_1 M_2)^{3/5}}{(M_1+M_2)^{1/5}}~,
\label{eq:chirp}
\ee
 is the so-called chirp mass of the system.
It is important to remark that GW detectors 
do not measure the source-frame chirp mass, but the redshifted mass $\widetilde{M}_c$, 
related to the source-frame mass by the relation $\widetilde{M}_c = (1+z)M_c$, where $z$ is the redshift of the merging system with respect to the detector.
The amplitude of a GW is proportional to the inverse of the luminosity distance $D_L = (1+z)D_{c}$, where the comoving distance is
\begin{equation}\label{eq:RedShift}
D_c =  \frac{c}{H_0}\int_0^z\frac{dt}{E(t)}~,~~~~~E(z) = \sqrt{\Omega_M(1+z)^3 + \Omega_k(1+z)^2 + \Omega_{\Lambda}}~,
\end{equation}
with $t_H \equiv 1/H_0 = 13.969$ Gyr the Hubble time.
For a given luminosity distance, and assuming standard cosmology (that corresponds to $\Omega_k = 0$, $\Omega_{\Lambda} = 0.7$, and $\Omega_{M} = 0.3$), it is possible to extract the redshift $z$, 
and convert the detector-frame chirp mass into the source frame chirp mass.

We can use the signal-to-noise ratio 
criterion $\rho \geqslant 8$ to estimate the sensitivity range of the LIGO detector in terms of mass and compactness.
For illustration, we take the limit of equal masses $M_1 = M_2 = M$.
Since we are limiting the analysis to the inspiral phase, \eq{eq:Strain} is valid in the frequency range $f < f_{\rm ISCO}$ and we cutoff the integral in \eq{intrho} at $f = f_{\rm ISCO}$.
\begin{figure}[!t!]
\centering
  \includegraphics[width=.5\linewidth]{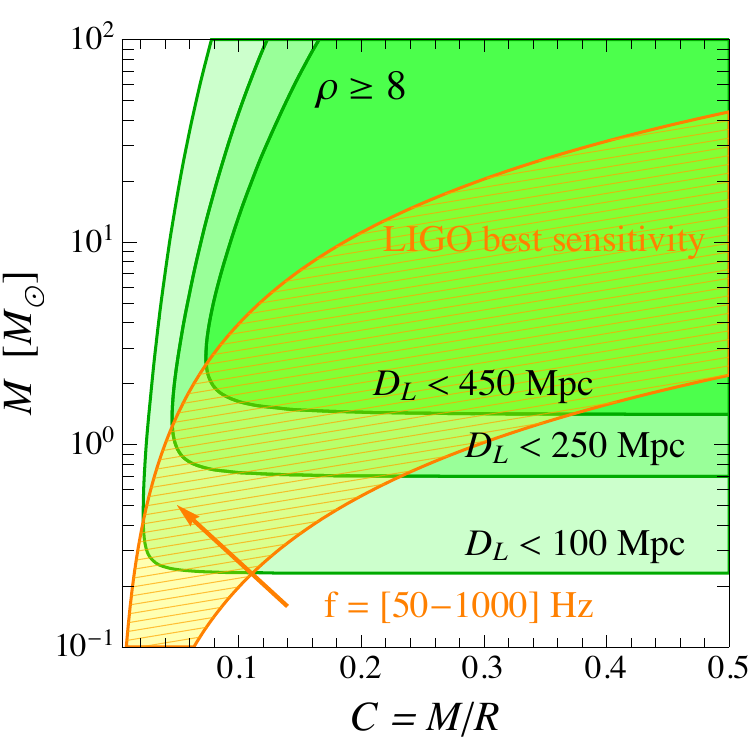}
\caption{\em 
The LIGO best sensitivity range in ECO mass $M$ and compactness $C$, for equal mass binary GW events. The yellow band corresponds to the GW frequency range $f=[50-1000]$~Hz, the green regions to a signal-to-noise ratio for an inspiral event occurring within the luminosity distance $D_L$, taking $\rho \geqslant 8$ as a criterion for detection.
 }
\label{fig:SNR}
\end{figure}
We show our result in \fig{fig:SNR}. The yellow band shows the LIGO frequency sensitivity range, given in \eq{eq:Range}. The green regions show the signal intensity sensitivity range (corresponding to $\rho \geqslant 8$), for a single GW event occurring within a distance $D_L$. We choose three particular values of the maximum luminosity distance, namely $D_L = 100,\,250,\,450$ Mpc.
Note that the corresponding redshift is very small, ranging from $z = 0.023$ for $D_L = 100$ Mpc to 
$z= 0.1$ for $D_L = 450$ Mpc. We therefore neglect the redshift effects in \fig{fig:SNR}. The overlap between the yellow and green regions identifies the LIGO best sensitivity in ECO mass $M$ and compactness $C$.

Let us now analyse in detail the implications of \eq{eq:Range} for the various models considered in \sect{sec:exo}.

\subsubsection{Boson Stars}
We start our discussion with the boson stars described in \sect{sec:exobs}, in the case of the potential in \eq{eq:BosonPotential}, taking $\lambda >0$ in order to have repulsive self-interaction. 
Repulsive self-interactions play an important role in in establishing an equilibrium configuration
 since they balance -- in addition to the quantum repulsive force generated by the Heisenberg uncertainty principle -- the
attractive pull of gravity.
Furthermore, having in mind the new boson as a dark matter candidate, the inclusion of self-interactions 
could resolve the puzzles arising at galactic scales in collisionless cold
dark matter (CCDM) simulations.

As discussed in \sect{sec:exodm},
CCDM problems are resolved without contradicting other astrophysical observables
 for a cross-section per unit DM mass in the range  $0.1~{\rm cm}^2/{\rm g} \lesssim \sigma/m_{B} \lesssim 1~{\rm cm}^2/{\rm g}$. 
Using $\sigma = \lambda^2/64\pi m_B^2$ we find~\cite{Eby:2015hsq}
\begin{equation}\label{eq:AllowedLambda}
\left(\frac{m_B}{{\rm MeV}}\right)^{3/2}\lesssim
\frac{\lambda}{10^{-3}}
\lesssim
3\times \left(\frac{m_B}{{\rm MeV}}\right)^{3/2}~.
\end{equation}
Following~\cite{Colpi:1986ye} (see also~\cite{Eby:2015hsq})  it is possible to describe the equilibrium configurations of the boson star system 
by numerically solving the coupled Einstein and Klein-Gordon equations.

\begin{figure}[!t!]
\minipage{0.5\textwidth}
  \includegraphics[width=.975\linewidth]{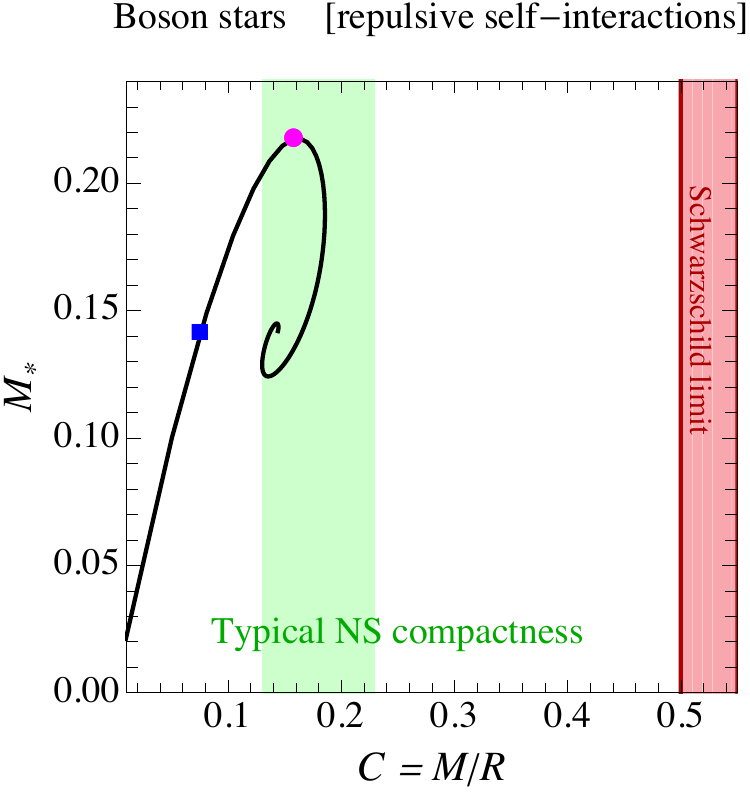}
\endminipage\hfill
\minipage{0.5\textwidth}
  \includegraphics[width=.95\linewidth]{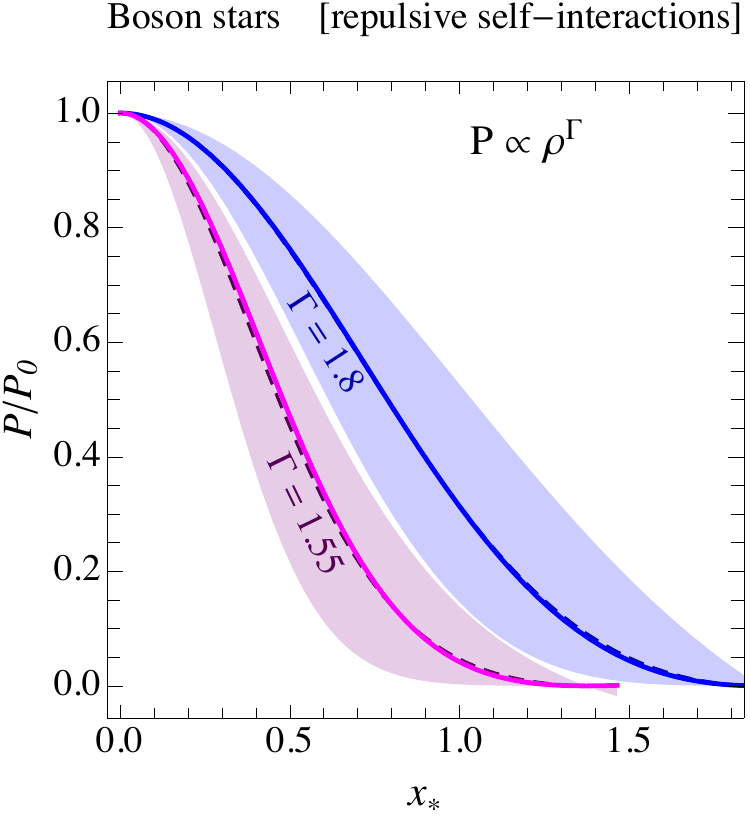}
\endminipage \vspace{.25 cm}
\caption{\em 
Left panel. Mass-compactness  relation in the case of boson stars with repulsive self-interactions.
The dimensionless mass $M_*$ is defined in \eq{eq:DimensionlessMass}.
The region shaded in red exceeds the compactness of a Schwarzschild BH, $C_{BH} = 1/2$.
The region shaded in green represents the typical NS compactness.
The magenta circle and blue square represent boson stars with pressure plotted in the right panel.
Right panel. Pressure (solid magenta and blue lines, normalised w.r.t. the value at the origin) as a function of the dimensionless radius $x_*$, defined as $x_* \equiv m_B^2 r/(M_{PL}\sqrt{\lambda/4\pi})$ (see~\cite{Eby:2015hsq,Colpi:1986ye}).
The blue and purple regions span the polytropic relation $P \propto \rho^{\Gamma}$ with $\Gamma = [1-3]$ (respectively, on the right- and left-most part of the band).
The dashed blue (magenta) line corresponds to $\Gamma = 1.8$ ($\Gamma = 1.55$).
}
\label{fig:PhaseDiagramBoson}
\end{figure}
These equilibrium solutions are characterised by 
 specific values of the mass $M$ and radius $R$. In the left panel of \fig{fig:PhaseDiagramBoson} we show the equilibrium solutions
after expressing $M$ and $R$ in terms of the compactness $C$ and  
the dimensionless mass $M_*$ of the system, defined as
\begin{equation}\label{eq:DimensionlessMass}
M_* =\left( \frac{M}{1.64 \times 10^6\,M_{\odot}}\right) \left( \frac{m_B}{\rm MeV}\right)^2\left( \frac{4\pi}{\lambda}\right)^{1/2} \, .
\end{equation}
The green band shows the typical NS compactness, while in the red region we have $C > C_{BH}$.
Note that the system is stable -- stability here refers to the response of the equilibrium solutions  under linear perturbations -- on the branch of the black line to the left of the turning point marked by a magenta dot in \fig{fig:PhaseDiagramBoson}.
Beyond the turning point, the equilibrium solutions become unstable.\footnote{Stability of boson stars was discussed in refs.~\cite{Lee:1988av,Gleiser:1988rq,Gleiser:1988ih} by 
means of
a mode analysis in linear perturbation theory. 
The general result is that transition from stability to instability always
occurs at critical points of the boson star mass $M$
 as a function of the value of the scalar  field at the origin (that is where $dM/d\phi_0=0$). Beyond the critical point, the boson star
  either collapse to a BH or disperse
to infinity. Other approaches based on fully numerical non-linear evolutions~\cite{Liebling:2012fv} and catastrophe theory~\cite{Kusmartsev:1990cr} confirmed this result.}
This also determines the maximum compactness for a self-interacting boson star, $C_{\rm max} \approx 0.158$, which is in agreement with the expectation given in \eq{bosstar3}.

Figure~\ref{fig:PhaseDiagramBoson} refers to the macroscopic properties of the boson stars in terms of dimensionless quantities.
Although this information is crucial to capture the scaling properties of the system, it is equally important -- in 
light of the limited detector sensitivity, and to provide a particle physics perspective -- to highlight the relation between the boson star mass and the microscopic 
parameters $m_B$, $\lambda$.

Recalling the definition in \eq{eq:DimensionlessMass} it is possible to 
express the LIGO sensitivity shown in \fig{fig:SNR}
in terms of the microscopic parameters $m_B$, $\lambda$ and of the mass $M$, while $C$ is derived from \fig{fig:PhaseDiagramBoson}.

\begin{figure}[!htb!]
\centering
  \includegraphics[width=.55\linewidth]{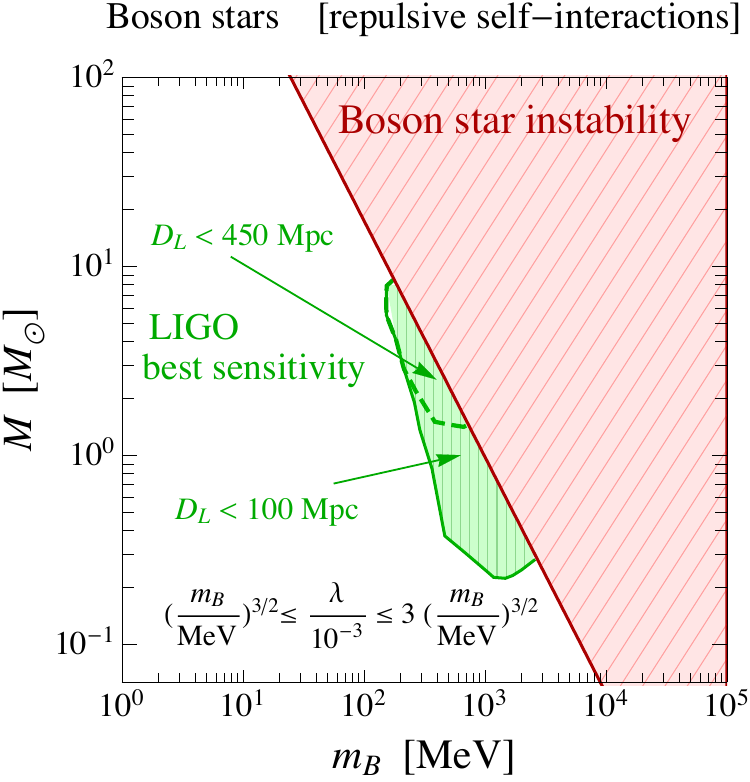}
\caption{\em 
LIGO best sensitivity (region shaded in green, defined according to \fig{fig:SNR} with $D_L = 450$ Mpc (dashed contour) and $D_L = 100$ Mpc (solid contour)) in terms
of  boson star mass $M$ and dark matter mass $m_B$. We restrict the analysis to self-couplings
 in the range given by \eq{eq:AllowedLambda} in order to make contact with the solution of the CCDM problems.
 The red region is excluded by the condition $M_* >  (M_{*})_{\rm max} \approx 0.22$.
}
\label{fig:BosonStarScan}
\end{figure}

For simplicity, we take the case of equal masses ($M_1 = M_2$). We scan over the parameters $m_B$, $\lambda$ and, in order to be consistent with the solution of the CCDM problems, 
we restrict the allowed values of $\lambda$ according to \eq{eq:AllowedLambda}).
We show our result in \fig{fig:BosonStarScan}, giving the LIGO sensitivity region for GW events within $D_L = 100$ or 450 Mpc.
The region shaded in red is excluded since the total mass exceeds the maximum value $(M_{*})_{\rm max} \approx 0.22$. Note that the boundary of the red region is in agreement with the parametric scaling given in \eq{bosstar2}.
The  green region represents the portion of parameter space in which 
the LIGO detector reaches the best sensitivity for GW events within a luminosity distance $D_L < 100$ Mpc (entire region) or $D_L < 450$ Mpc (region limited by the dashed line). Boson stars with mass in the range $M \approx [1-20]~M_{\odot}$ fall within the detector sensitivity. If boson stars are sufficiently numerous in the universe and GW events are produced within distances $D_L < 100$ Mpc, then the sensitivity range extends to $M \approx [0.2-20]~M_{\odot}$
Interestingly, the analysis points towards a specific and narrow range of values for the dark particle mass, that is $m_B \approx [60\,{\rm MeV}-3\,{\rm GeV}]$. 

This simple exercise shows that it is possible to construct boson stars without violating existing astrophysical constraints, 
favoured by the solution of the CCDM problems, 
well within the sensitivity of the LIGO detector, and with masses also in the gap between NS and BH.

The LIGO sensitivity range of dark particle masses shown in \fig{fig:BosonStarScan} does not correspond to the typical expectation for a WIMP, a dark matter candidate whose present density is determined by thermal freeze-out. On the other hand, it is particularly interesting for asymmetric dark matter (for a review, see~\cite{Zurek:2013wia}). This class of models finds its justification in the observation that the amount of dark matter and baryonic matter in the universe is of the same order of magnitude. Therefore, a new stable particle with mass and cosmic asymmetry comparable with the baryonic case can naturally explain the observed dark matter. More quantitatively, one finds that the asymmetric dark matter particle must have a mass
\beq
m_{DM} = \frac{\eta_b}{\eta_{DM}}\ 5\ \gev \, ,
\eeq
where $\eta_{DM,b}$ are the dark-matter and baryon cosmic asymmetries, respectively. The ratio $\eta_{DM}/\eta_b$ is very model dependent. A non-vanishing asymmetry $\eta_{DM}$ is possible whenever the dark particle carries a conserved quantum number, which can happen both for a spin-0 boson (as considered here) and for a spin-1/2 fermion (as discussed in the following). Moreover, an interesting situation occurs when interactions in the high-energy domain violate baryon number and dark-matter number, but preserve a linear combination. Then, at primordial times, any pre-existing asymmetry is reshuffled between the two sectors and the ratio $\eta_{DM}/\eta_b$ 
is predicted to be of order unity (if the DM is relativistic at the time of asymmetry sharing), at least within an order of magnitude or so. For instance, this is the case of leptogenesis, in which electroweak sphalerons are expected to redistribute among baryons and leptons any original cosmic asymmetry induced by right-handed neutrino decays. Therefore, the LIGO sensitivity range in $m_B$ shown in \fig{fig:BosonStarScan} looks promising for the exploration of asymmetric dark matter models.

From the particle-physics point of view, the interpretation of the light bosonic particle as a realistic dark matter candidate raises a concern with the concept of naturalness. The most straightforward solution to the problem is to identify the new particle with a pseudo-Goldstone boson, in which a small parameter controls the breaking of the shift symmetry and generates the repulsive quartic self-interaction in the scalar potential. For instance, this is done in the construction of models in which the Higgs is a composite particle remaining light because of an approximate Goldstone symmetry (for reviews, see~\cite{Contino:2010rs,Panico:2015jxa}). 

An obstruction in having a repulsive quartic self-interaction was recently pointed out in~\cite{Fan:2016rda}, in the context of axion-like particles. Consider a Goldstone boson generated by the spontaneous breaking of a compact group and add a scalar potential that breaks the continuos shift symmetry, but preserve both a discrete subgroup and CP. (This is the case of the QCD axion, in which the PQ symmetry is explicitly broken by non-perturbative instanton effects.) Because of CP, the potential must contain only even powers of the field and, because of the discrete periodicity, the signs of the quadratic and quartic terms must be opposite. Then, the condition that the particle is non-tachyonic implies that the self-interaction must be attractive ($\lambda <0$). Nonetheless, it is possible to construct more involved models of axion-like particles with repulsive interactions~\cite{Fan:2016rda}. 

\medskip

To make the parallel with NS more effective, we complete our discussion by considering the equation of state.
NS can be modelled by polytropes with equation of state $P = K\rho^{\Gamma}$, where $K$ is a proportionality constant while the polytropic index takes the values
 $\Gamma = [2-3]$. In our analysis we derive the equation of state numerically, and we show our results in the right panel of \fig{fig:PhaseDiagramBoson}
 where we focus on two specific equilibrium solutions (magenta circle and blue square in the left panel of \fig{fig:PhaseDiagramBoson}).
 We plot the pressure (normalised w.r.t.\ the value at the origin) as a function of the dimensionless radial distance $x_*$.
 We find that the analysed  boson stars 
 are well described by a polytrope with $\Gamma = 1.55$ and $\Gamma = 1.8$.  Boson stars are therefore characterised by a softer equation of state
 if compared with typical NS. 
 
\begin{figure}[!htb!]
\minipage{0.5\textwidth}
  \includegraphics[width=.95\linewidth]{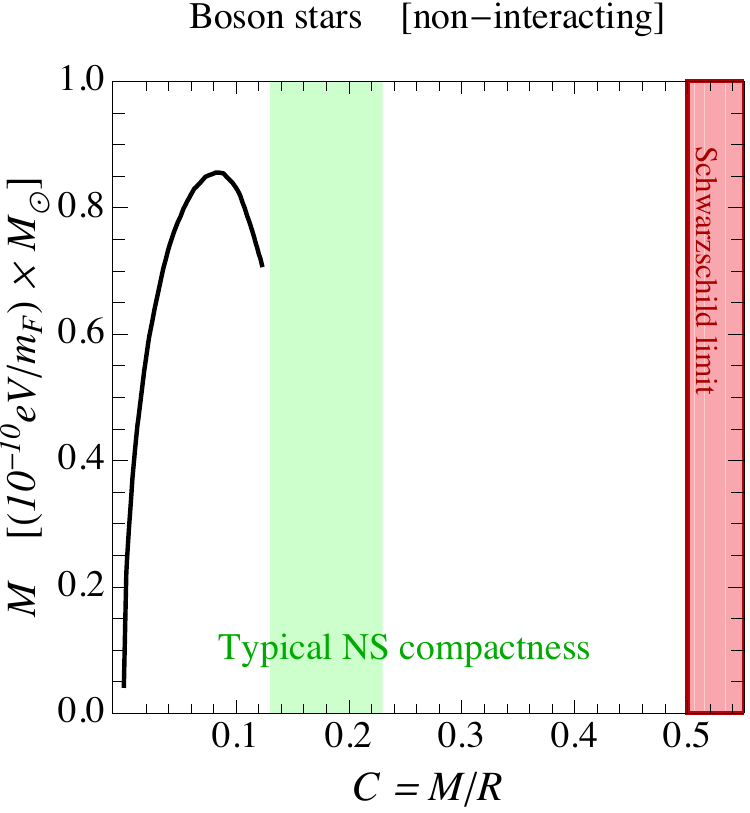}
\endminipage\hfill
\minipage{0.5\textwidth}
  \includegraphics[width=1\linewidth]{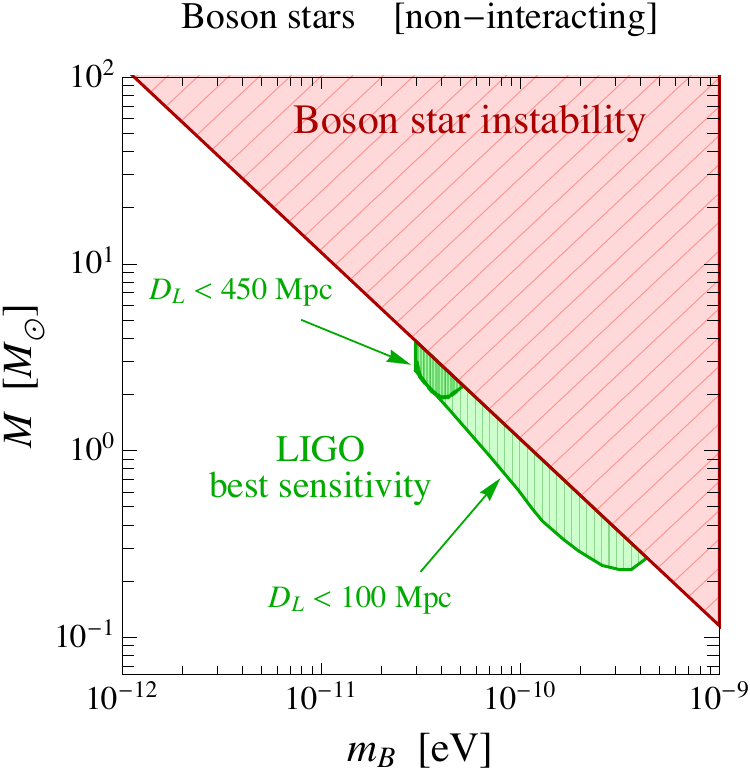}
\endminipage \vspace{.25 cm}
\caption{\em 
Left panel. Mass-compactness  relation in the case of boson stars without self-interactions.
Right panel.
LIGO best sensitivity (defined as in \fig{fig:SNR}) in terms of the boson star mass $M$ and dark matter mass $m_B$. 
}
\label{fig:FreeBosonStar}
\end{figure}

Before discussing fermion stars, let us comment about the case of boson stars without self-interactions
in which the equilibrium against gravitational collapse is entirely due to quantum effects.
In the left panel of \fig{fig:FreeBosonStar} we show the mass-compactness  relation
compared with the typical NS compactness and the BH limit. Note that 
the numerical analysis confirms the maximum mass  quoted in \eq{bosstar}.
Furthermore, as anticipated in \eq{bosstar4}, we find the corresponding maximum value of  
compactness $C_{\rm max} \approx 0.08$, for stable solutions to the left of the turning point.
In the right panel of \fig{fig:FreeBosonStar} we 
present the LIGO best sensitivity region 
in terms
of the boson star mass $M$ and the microscopic particle mass $m_B$.
 We implemented the bound in \fig{fig:SNR} 
 using a luminosity distance $D_L$ within 100 (lighter green) or 450 (darker green) Mpc.
Smaller luminosity distances allow for enlarging the green region 
 towards smaller values of the total mass, which fall more easily inside the LIGO range, as a consequence of the limited compactness. 
 Nevertheless, note that values as large as $M \simeq 4$ $M_{\odot}$ are possible; according to the left panel of \fig{fig:FreeBosonStar}, they correspond 
 to boson stars with compactness close to the maximum value $C_{\rm max} \approx 0.08$, which could be detected even assuming 
a luminosity distance as large as $D_L \simeq  450$ Mpc.

As shown in \fig{fig:FreeBosonStar}, for the case of the non-interacting boson LIGO is sensitive to the mass range $m_B \approx  [0.2-0.8]\times 10^{-10}$ eV, which can be extended to $m_B \approx  [0.2-4]\times 10^{-10}$ eV if there is enough probability of observing GW events within 100 Mpc. This range is  
quite different from the one of the interacting case because, for the free particle, the pull of gravity, which becomes stronger with $m_B$, 
is balanced only by  the quantum effects generated by the Heisenberg uncertainty principle. 

The LIGO sensitivity mass range for the non-interacting boson is interesting for axion-like particle interpretations. For the QCD axion, one finds
\beq
m_a = \left( \frac{10^{17}\ \gev}{f_a} \right) \ 0.6 \times 10^{-10}\ {\rm eV} \,
\eeq
where $f_a$ is the axion decay constant. Hence, LIGO can be sensitive to axion models in which the symmetry-breaking scale is in the range $1\times 10^{16}\ \gev \lesssim f_a \lesssim 3\times 10^{17}\ \gev$. This is an intriguing range of scales, since these are the values expected in GUT or String models. A QCD axion with decay constant in the GUT range can be a good dark matter candidate only if its field configuration at the end of inflation is sufficiently close to the minimum of its potential (which, in the absence of a dynamical explanation, corresponds to tuned initial conditions)~\cite{Sikivie:2006ni,Hertzberg:2008wr}. Moreover, it can be consistent with WMAP constraints on isocurvature fluctuations only if the Hubble constant during inflation is smaller than $10^9$--$10^{10}$~GeV. Therefore, detection of a non-vanishing tensor-to-scalar ratio in the cosmic microwave background could rule out a QCD axion with GUT-scale decay constant as dark matter candidate~\cite{Visinelli:2014twa}.

Finally, the new boson could be an axion-like particle, unrelated to the strong CP problem. In this case, its mass is expected to be $m_B \approx \Lambda^2/f_B$, where $f_B$ is the scale of spontaneous symmetry breaking and $\Lambda$ is the dynamical scale associated with the explicit breaking of the global symmetry. The LIGO sensitivity range for $m_B$ can be attained for a variety of new-physics scale. The boson particle can also play the role of dark matter, but without specifying the model we cannot make definite predictions.

\subsubsection{Fermion Stars}

\begin{figure}[!htb!]
\minipage{0.5\textwidth}
  \includegraphics[width=.975\linewidth]{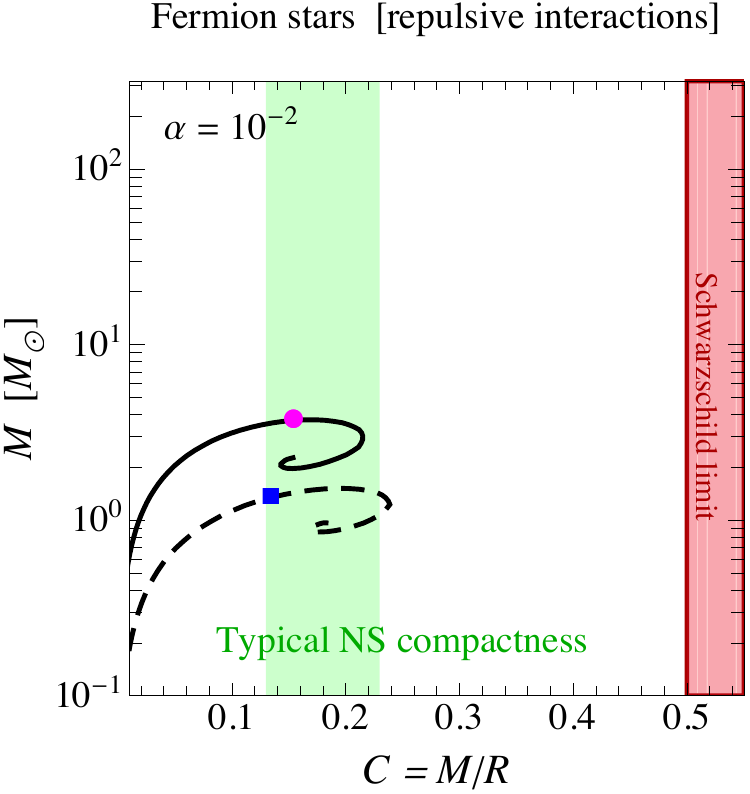}
\endminipage\hfill
\minipage{0.5\textwidth}
  \includegraphics[width=.95\linewidth]{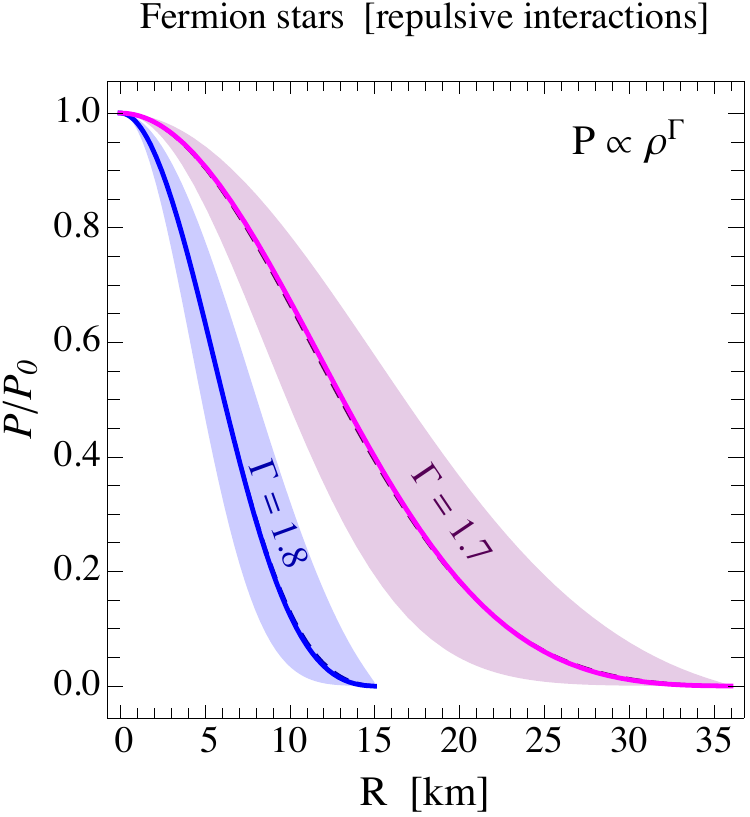}
\endminipage \vspace{.25 cm}
\caption{\em 
Left panel. Mass-compactness  relation in the case of asymmetric dark matter stars. The strength of the attractive interaction is $\alpha = 10^{-2}$, and 
we take the mediator mass $\mu = 50$ MeV. The dashed (solid) black line corresponds to 
$m_{F} = 1$ GeV ($m_{F} = 500$ MeV).
The region shadowed in red exceeds the compactness of a Schwarzschild BH, $C = 1/2$.
The region shadowed in green represents the typical NS compactness.
The magenta circle and blue square represent boson stars with pressure plotted in the right panel.
Right panel. Pressure (solid magenta and blue lines, normalised w.r.t. the value at the origin) as a function of the radius $R$.
The blue and purple regions span the polytropic relation $P \propto \rho^{\Gamma}$ with $\Gamma = [1-3]$ (respectively, on the right- and left-most part of the band).
The dashed blue (magenta) line corresponds to $\Gamma = 1.8$ ($\Gamma = 1.7$).
}
\label{fig:PhaseDiagramFermion}
\end{figure}

Let us now move to discuss the case of fermion star.
We follow~\cite{Kouvaris:2015rea}, and focus on the case of fermionic dark matter $\chi$ with mass $m_F$ and repulsive interactions. The latter are parameterised 
by a Lagrangian term of the form $g\phi_{\mu}\overline{\chi}\gamma^{\mu}\chi$, where $\phi_{\mu}$ is a vector mediator with mass $\mu$ and $g$ the coupling constant.
As for boson stars, the presence of self-interactions may play an important role in solving the CCDM problems.   
The equilibrium configurations can be obtained by solving the Oppenheimer-Volkoff  equation.
In the left panel of \fig{fig:PhaseDiagramFermion} we show 
two particular equilibrium solutions 
where we fixed $\alpha \equiv g^2/4\pi = 10^{-2}$, $\mu = 50$ MeV and $m_F = 500$ MeV (black solid line),  $m_F = 1$ GeV (black dashed line).
Note that at this stage we limited the analysis to specific values of masses and coupling since in full generality 
it is not possible, unlike the case of boson stars with self-interactions, to extract dimensionless scaling properties. 
The maximum allowed value for the fermion star mass $M$ is found from the turning point in the mass-central density relation; 
this is shown in the inset plot in the left panel of \fig{fig:PhaseDiagramFermion} for the case with $m_F = 500$ MeV.

In the right panel of \fig{fig:PhaseDiagramFermion} we analyse -- as done for boson stars -- the equation of state,
focusing our attention on two specific equilibrium solutions (magenta circle and blue square in the left panel of \fig{fig:PhaseDiagramFermion}).
 We find that the analysed  fermions stars 
 are well described by a polytrope with $\Gamma = 1.7$ and $\Gamma = 1.8$.
 
\begin{figure}[!htb!]
\minipage{0.5\textwidth}
  \includegraphics[width=.975\linewidth]{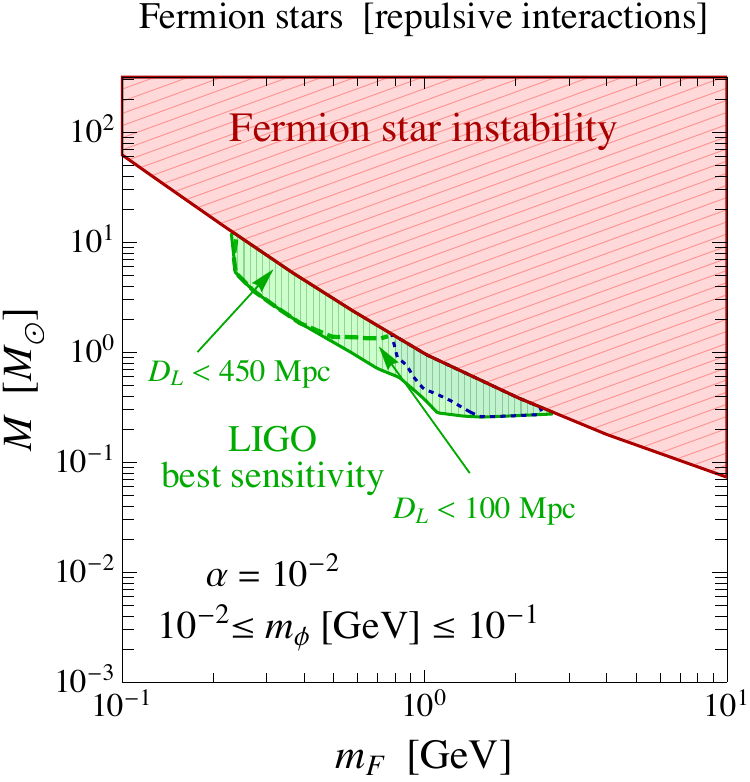}
\endminipage\hfill
\minipage{0.5\textwidth}
  \includegraphics[width=.95\linewidth]{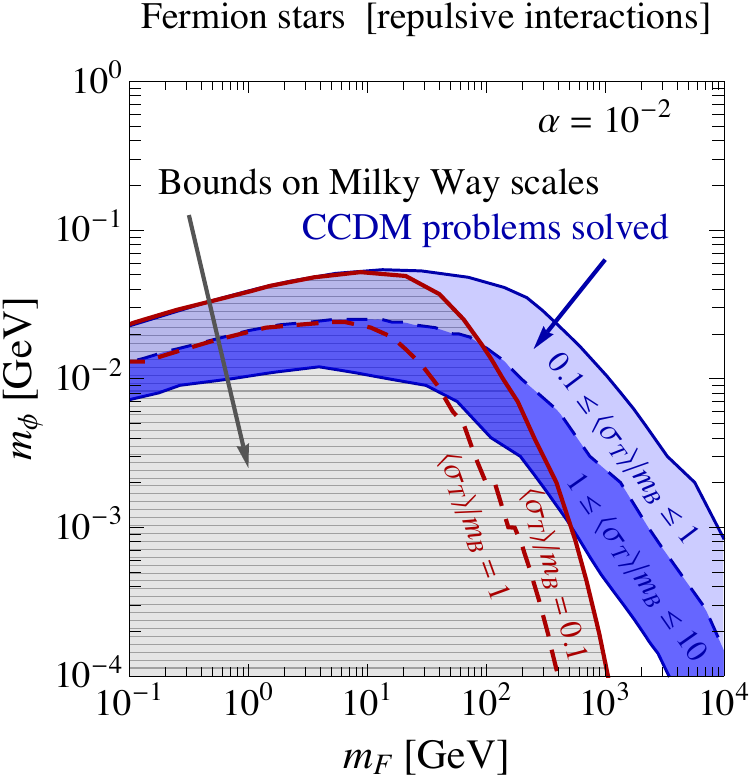}
\endminipage \vspace{.25 cm}
\caption{\em 
Left panel.
LIGO best sensitivity (region shaded in green, defined according to \fig{fig:SNR} with $D_L = 450$ Mpc (dashed contour) and $D_L = 100$ Mpc (solid contour)) in terms
of  fermion star mass $M$ and dark matter mass $m_F$. 
We restrict the analysis to mediator masses in the range $m_{\phi} = [10^{-2}-10^{-1}]$ GeV.
 The red region is excluded by the condition $M >  M_{\rm max}$, which we impose at each point in our sampling of the parameter space.
 The blue dotted line delimits the region in which $\langle\sigma_T \rangle /m_F = [0.1-1]$  cm$^2$/g for a dwarf halo with $v_0 = 10$ km/s
 and $\langle\sigma_T \rangle /m_F  < 0.1$  cm$^2$/g for a Milky Way-like halo with $v_0 = 200$ km/s.
 The coupling constant is fixed at $\alpha = 10^{-2}$.
 Right panel. Parameter space consistent with astrophysical observation. See text for a complete explanation.
}
\label{fig:FermionStarScan}
\end{figure}

To offer a broader perspective on the microscopic values of masses and coupling favoured by the LIGO sensitivity and compatible with astrophysical constraints, 
we present in \fig{fig:FermionStarScan} the result of a parameter scan.
As far as the impact of astrophysical constraints is concerned, we follow~\cite{Tulin:2013teo}.
As already anticipated in \sect{sec:exo}, 
dark matter self-interactions have important consequences on structure formation -- from dwarf galaxies to galaxy clusters.
This is especially true -- contrary to the case of boson stars in which we considered a contact self-interaction -- in the presence of non-trivial velocity and angular dependence of the scattering cross section. In this case the 
relevant quantity constrained by astrophysical observations is the velocity-averaged transfer cross section~\cite{Tulin:2013teo}
\begin{equation}
\langle\sigma_T \rangle = \int \frac{d^3 v}{(2\pi v_0^2)^{3/2}} e^{-\frac{1}{2}v^2/v_0^2}\sigma_T(v)~,
\end{equation}
where $v_0$ is the characteristic velocity dispersion of a given galactic structure (that is $v_0 \approx 10$ km/s for a dwarf galaxy and 
$v_0 \approx 200$ km/s for a Milky-Way--like elliptic galaxy) and $v$ is the relative velocity between the two dark matter particles participating in the scattering process.
The transfer cross section $\sigma_T$ is defined by
\begin{equation}
\sigma_T = \int d\Omega (1-\cos\theta)\frac{d\sigma}{d\Omega}~,
\end{equation}
where $\theta$ is the scattering angle.

As discussed in eq.~(\ref{eq:GalaxyCrossSection}) the most stringent bound comes from the observed ellipticity of Milky-Way--sized haloes which imposes the condition
 $\langle\sigma_T \rangle /m_F \lesssim 0.1-1$ cm$^2$/g for $v_0 \approx 200$ km/s. 
 In the right panel of \fig{fig:FermionStarScan} we mark in grey (with horizontal meshes)  the corresponding excluded region in the parameter space 
 $(m_F, m_{\phi})$ with fixed $\alpha= 10^{-2}$. The solid (dashed) line refers to the more (less) stringent bound  
 $\langle\sigma_T \rangle /m_F = 1$ cm$^2$/g (or $0.1$ cm$^2$/g).
 
A velocity-averaged transfer cross section in the region 
$\langle\sigma_T \rangle /m_F = [0.1-10]$  cm$^2$/g for $v_0 \approx 10$ km/s is required to solve to CCDM problems in dwarf 
galaxies, see eq.~(\ref{eq:DwarfCrossSection}).
 In the right panel of \fig{fig:FermionStarScan} we mark in blue 
 the corresponding region. The darker area restricts the allowed range to $\langle\sigma_T \rangle /m_F = [1-10]$  cm$^2$/g.
 
Solutions to the CCDM problems that are clearly not in tension with the constrains from Milky Way-like haloes 
are found only for values $m_F \gtrsim 10^2$ GeV.
For smaller dark-matter mass,  in particular below  a few GeV, solutions are possible only assuming the less stringent bound in Milky Way-like haloes
$\langle\sigma_T \rangle /m_F \lesssim 0.1$ cm$^2$/g.
 
 Equipped with these results, 
 we are now in the position to study the interplay with LIGO. In the left panel of \fig{fig:FermionStarScan} we show 
the LIGO best sensitivity region (shaded in green) in the parameter space $(M, m_F)$.
We scanned over the mediator mass in the range $m_{\phi} = [10^{-2}-10^{-1}]$ GeV.

For each point, we computed the maximum mass  allowed by 
the solution of the Oppenheimer-Volkoff  equation (see caption in \fig{fig:PhaseDiagramFermion} and related discussion);
the envelope of these points defines the boundary of the red region shown in \fig{fig:FermionStarScan}, which corresponds to star instability.

The region limited by the blue dotted line encases the points with 
$\langle\sigma_T \rangle /m_F = [0.1-1]$  cm$^2$/g 
that are compatible with the solution of the CCDM problems and not excluded by observations in 
Milky-Way--sized haloes if the less stringent bound $\langle\sigma_T \rangle /m_F \lesssim 0.1$ cm$^2$/g
is imposed. Note that the region with $m_F \simeq [10^2-10^3]$ GeV, $m_{\phi} = [10^{-2}-10^{-1}]$ GeV
in which the CCDM problems can be solved without any tension 
with astrophysical observations would correspond to fermion stars with small mass, far away from the LIGO sensitivity.

Our scan shows that, in the mass range $m_F = [0.1-2]$ GeV, 
it is possible to construct a fermion star consistent with astrophysical observations. As previously discussed in the case of boson stars, this range of values for $m_F$ is especially promising in the context of asymmetric dark matter.
Furthermore, besides  the mass coincidence, another important aspect of the model described in this section, in the context of asymmetric dark matter, is the presence of a lighter massive vector mediator. In fact, the Lagrangian of the simplest model of asymmetric dark matter is based on 
 a  dark  Abelian  gauge group spontaneously broken, 
 and the condition $m_F > m_{\phi}$ -- with $\phi$ the dark photon --  is important since it opens up the annihilation channel $\overline{\chi}\chi \to \phi\phi$,
which is crucial to suppress
the symmetric component of the relic density~\cite{Kaplan:2009ag}.

\subsection{Gravitational Waveforms from ECO Mergers}\label{sec:waveforms}
\begin{figure}[t]
\begin{center}
\includegraphics[width=0.6\textwidth]{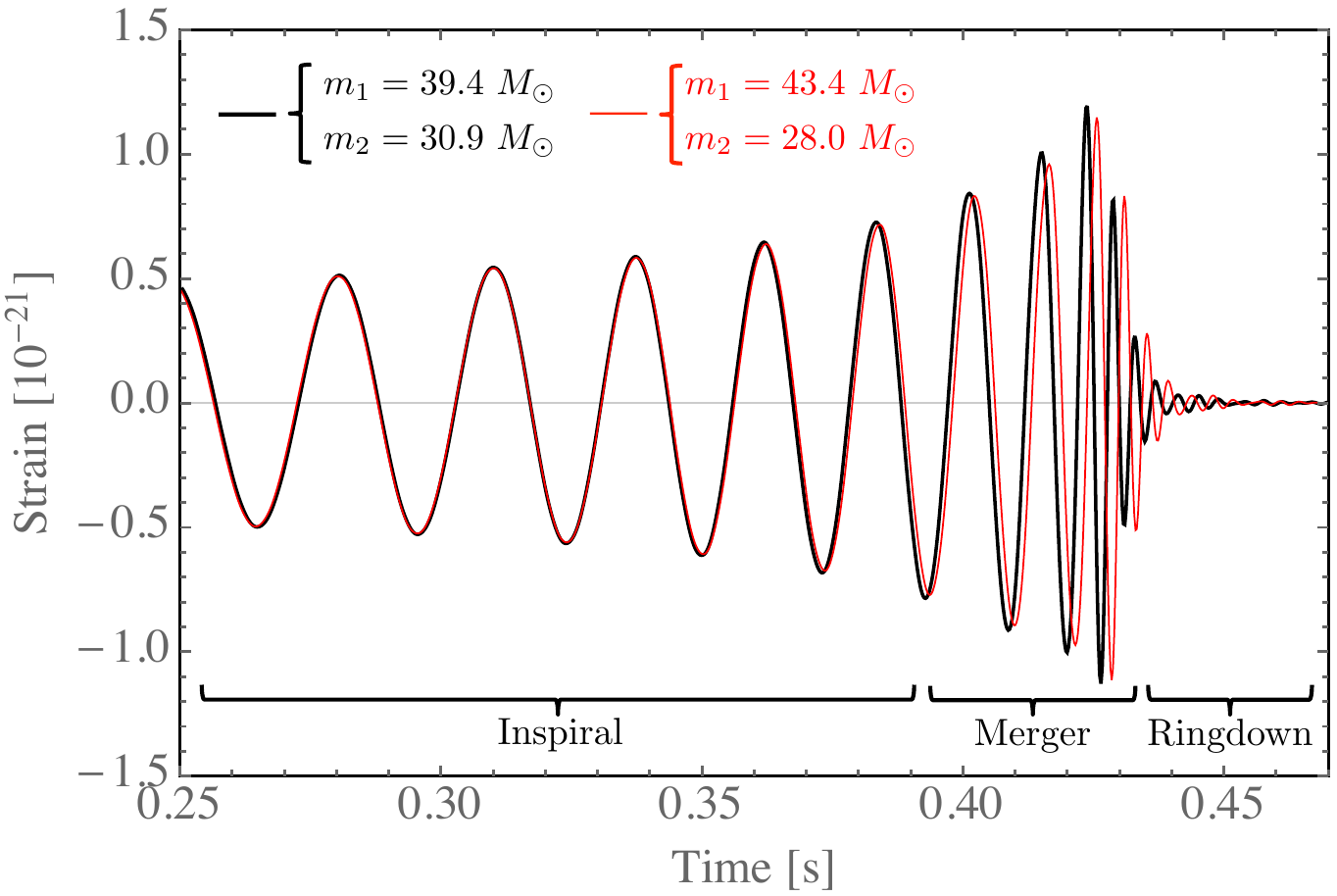}
\caption{\em The three phases of a black hole binary merger for mass parameters providing the best-fit to the LIGO detection (black) and mass parameters which give the same chirp mass, but are excluded at $90\%$ CL for a mass-ratio parameter of $q=0.65$ (red).  Mass parameters are taken from the EOBNR fit of \cite{TheLIGOScientific:2016wfe} and waveforms are calculated using the analytic formulae of \cite{Ajith:2009bn}.  As can be seen, the inspiral phase can be virtually identical for both choices of parameters, with the significant differences only arising in the later stages of the merger.  Even then, these differences seem relatively small, demonstrating the sensitivity of LIGO to modified waveforms.
\label{fig:GW150914}}
\end{center}
\end{figure}

As we expect that all astrophysical objects of mass $M \gtrsim 5 M_\odot$ are black holes, and we know that the nature of the gravitational waveform for BH-BH merger events is rather specific, evidence for the existence of ECOs may arise from a single observation of a merger event if the observed waveform is sufficiently exotic.  In fact, if the amplitude is comparable to the recently observed event GW150914 then the waveform need not be radically different in order to obtain evidence of new physics.  To illustrate this fact, using GW150914 as an example, we may consider the sensitivity of LIGO by comparing the waveform for the best-fit parameters with the waveform for a choice of primary mass parameters which is excluded at 90\% CL by the LIGO collaboration from a likelihood analysis.  In \fig{fig:GW150914} we show in black a waveform generated using the analytic approximation of \cite{Ajith:2009bn} for the best-fit mass parameters determined by the LIGO collaboration \cite{TheLIGOScientific:2016wfe}.  In solid red we show an analogous waveform for two masses which reproduce exactly the best-fit chirp mass but with a mass ratio $q=M_1/M_2=0.65$, excluded at 90\% CL~\cite{TheLIGOScientific:2016wfe}.  In \fig{fig:GW150914} we see that the differences in the two waveforms are very small, demonstrating that, for an event of amplitude comparable to GW150914, LIGO is already sensitive to relatively minor modifications of the waveform in the merger and ringdown phases.  This suggests that even small departures from the BH-BH waveform may be sufficient to give convincing evidence of new physics.

\subsubsection{Finite size effects}
Strong modifications of the gravitational waveform may arise if an ECO of mass $M$ has a radius significantly larger than the Schwarzschild radius or, in other words, a compactness parameter $C = M/R$ much smaller than the Schwarzschild value $C_{BH}=1/2$.    The merger phase will begin earlier than expected for black holes of the same mass.  We may use \eq{eq:Freq2} to estimate how this effect manifests.  As the ISCO frequency characterises the onset of the merger phase we consider the ratio of the ISCO frequency for an object of compactness $C$ relative to a BH of the same mass
\beq
 \frac{f^{\text{ISCO}}_{BH}}{f^{\text{ISCO}}_{ECO}}  =   (2 C)^{-3/2} \approx 5.5 \left(\frac{0.16}{C}\right)^{3/2} ~.
\eeq

This quantity is useful as it implies that at the onset of the merger phase the wavelength of GW emission in a binary system of boson or fermion stars will be a factor of $5.5\times (0.16/C)^{3/2}$ longer than for a BH-BH merger.  Essentially, the onset of the merger is earlier and the highest frequencies of the inspiral phase will not be reached.  This crude qualitative feature can be made more explicit by considering the full waveform.

The calculation of the waveform modifications is a complex task, involving numerical GR and hydrodynamics.  Let us first consider ECOs supported by fermion degeneracy pressure (ECO$_F$), as discussed in \sect{sec:exo}.  The physics of these ECOs is in many ways analogous to the physics of neutron stars (NS), which will serve as our guide.  Studies of GWs from NS-NS mergers have been performed and we will consider the results of \cite{Faber:2002cg}.  The details of the waveform depend on the mass-radius relationship and the EoS, which in turn depends on a number of factors including whether or not the fermions are relativistic.  We will consider an EoS described by $\Gamma = 2$ as this is typical for a NS.

\begin{figure}[t]
\begin{center}
\includegraphics[width=0.98\textwidth]{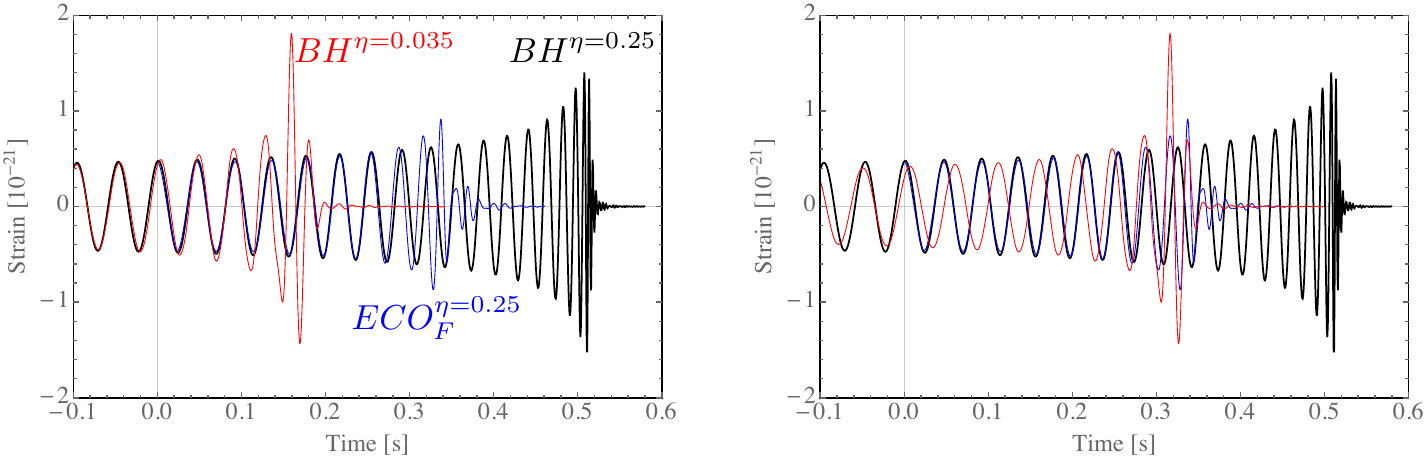}
\caption{\em An example comparison between the waveform for a BH-BH merger with equal mass ($35 M_\odot$) primary BHs (black), versus an approximation of an ECO$_F$-ECO$_F$ merger of two ECO$_F$s of the same mass (blue), and a BH-BH merger chosen to have the same chirp mass, but with a large mass ratio (red), attempting to mimic the ECO$_F$ signature.  The only difference between panels is the position of the red line, chosen on the left for comparison of inspiral phases, and on the right for comparison of the ISCO frequency.  The blue waveform is an NS-NS result from \cite{Faber:2002cg}, rescaled to the same primary masses as in the BH-BH merger.  This is used as a guide, because an ECO$_F$-ECO$_F$ merger would have modifications of a similar form.  The comparison demonstrates that although the inspiral phase may look similar to a BH-BH merger, the onset of the merger phase occurs earlier and the merger and ringdown phases are significantly different, offering the possibility of discovering evidence for for the existence of a massive ECO$_F$ through a single GW observation.  The only known objects with compactness in this range are NSs, which are much lighter, such that an ECO$_F$ merger in the mass range $\gtrsim M_\odot$ would not be mistaken for an NS merger.
\label{fig:NSNS}}
\end{center}
\end{figure}

In \fig{fig:NSNS} we show again a GW signature similar to GW150914 in black and compare it to the waveform for a merger of two ECO$_F$ in blue.  The latter was extracted from the waveform of an NS-NS merger calculated in \cite{Faber:2002cg}.  In \cite{Faber:2002cg} the two masses were taken as $M_{NS} = 1.4 \, M_\odot$ and the compactness as $C=0.16$.  Technically, if the mass-radius ratio is the same, as it would be for a more massive ECO$_F$, the specific waveform is also valid for higher mass systems if the time and amplitude are rescaled proportional to the new total mass.  Thus, for comparison, we have rescaled the waveform to have primary masses equal to the BH-BH waveform, hence giving a frequency and amplitude in the inspiral phase equal to the BH-BH merger.  As can be clearly seen, even if the inspiral phase looks very similar to the signature expected of a BH-BH inspiral, as the wave evolves significant differences from a BH-BH merger may arise.  The onset of the merger phase occurs earlier and, by comparing the wavelength of GW emission in the ECO$_F$-ECO$_F$ case against the BH-BH case we estimate a wavelength ratio of $\sim 3.1$ at the onset of the merger phase, which is not too far away from our earlier simplistic estimate of $\sim 5.5$.  After the merger the remainder of the waveform is also significantly different. Of course, the final discrimination power depends on the precision of the experimental measurement of the waveform. For a related discussion, see ref.~\cite{Sampson:2014qqa}. 

Would it be possible for a BH merger to mimic the signature of an ECO$_F$ merger? To answer this question, let us express $f^{\text{ISCO}}_{BH}$ in terms of the chirp mass $M_c$ and the symmetric mass ratio
\be\label{eq:SymRatio}
\eta = \frac{M_1 M_2}{(M_1 +M_2)^2} ~.
\ee
Using $M_c = M_{\rm tot}\eta^{3/5}$, we find
\begin{equation}\label{eq:Freqeta}
f^{\rm ISCO}_{BH} = \frac{(1+\Delta_{BH} )\eta_{BH}^{3/5}}{ 6^{3/2} \pi M_{c}} ~,~~~~~
f^{\rm ISCO}_{ECO} = \frac{(1+\Delta_{ECO} ) \eta_{ECO}^{3/5}C^{3/2}}{ 3^{3/2} \pi M_{c}} ~ .
\end{equation}
Superficially, it may seem that BH mergers can always mimic the effect of compactness, since with an appropriate choice of $\eta_{BH}$ one can reproduce the same chirp mass and the same $f^{\rm ISCO}$ of an ECO event. However, other features of the waveform, such as the amplitude and its variation, can be used to distinguish the two cases, as long as $C$ is sufficiently smaller than $1/2$. 

To illustrate this point in \fig{fig:NSNS} we show the waveform for a BH merger in which the two BH masses have been chosen to reproduce the same chirp mass and ISCO frequency as the ECO$_F$ merger.\footnote{From \eq{eq:Freqeta} we find that a value $\eta_{BH}=0.035$ is required to reproduce the ISCO frequency of the ECO$_F$.  Strictly speaking, this value is a factor $\sim2$ below the lower limit of the range of validity for the analytic expressions, as discussed in \cite{Ajith:2009bn}.  However, as also discussed in \cite{Ajith:2009bn}, for non-spinning BHs the analytic approximation should reproduce the test mass limit, thus $\eta_{BH}=0.035$ is most likely still acceptable for $S_1=S_2=0$.  Interestingly, we see in the right panel of \fig{fig:NSNS} that the estimate of the ISCO frequency from \eq{eq:Freqeta} does indeed match the ECO$_F$ ISCO frequency well.}  It is clear from \fig{fig:NSNS} that the BH merger and ringdown waveform is still sufficiently different from the ECO$_F$ waveform that it would be possible to discriminate the two for the compactness motivated by the particle physics models of \sect{sec:conv}, even in the case of similar chirp mass and ISCO frequency.

In addition, we argue that it is very unlikely for a BH system to mimic the same chirp mass and ISCO frequency of an ECO in the first place. From \eq{eq:Freqeta}, we see that this happens whenever
\be
\eta_{BH} = \eta_{ECO} \left(\frac{1+\Delta_{ECO}}{1+\Delta_{BH}} \right)^{5/3} \left( 2 C \right)^{5/2}\ .
\ee
Considering the inequality $\eta_{ECO} \leq 1/4$ and the extremal values of the post-Newtonian correction $\Delta$ for non-spinning BHs, this corresponds to
\be
\eta_{BH} \leq 0.09 \times \left( \frac{C}{0.2} \right)^{5/2} ~.
\label{eq:unlikely}
\ee
As we will see when we come to discuss the mass distribution of BH binary systems in \sect{sec:census}, values of $\eta_{BH}$ satisfying \eq{eq:unlikely} for $C\lsim 0.3$ are highly unlikely to be realised in nature, as shown in \fig{fig:TestMass}.  

To summarise, a BH merger is unlikely to mimic the signature of an ECO, at least for the sufficiently small $C$ values motivated in ECO scenarios motivated by the specific BSM particle physics models discussed in \sect{sec:conv}. This is because the necessary mass ratio required to reproduce both the chirp mass and the merger frequency is typically too extreme to be likely to occur in the universe, according to stellar BH formation models. Moreover, the waveform is sufficiently distinct to allow determination of whether it originated from an ECO merger, as long as $C$ is sufficiently different than the BH value. This is illustrated by the comparison of \fig{fig:NSNS} with \fig{fig:GW150914}, which shows that the differences in waveform would be within the resolution of LIGO.  By observing such significant deviations in the waveform for a binary system involving large comparable mass objects, evidence for new physics may emerge.\footnote{Technically a BH-NS system may give rise to a chirp mass as large as expected for a BH-BH system, however this would require an extremely massive BH.}

As pointed out in~\cite{Pani:2009fd,Pani:2010jz}, the information on the spin of the merging objects, which can be extracted from the inspiral phase, can give further hints about the BH vs ECO nature of the binary source. This is because very compact objects without an event horizon (such as ECOs) develop a strong ergoregion instability when spinning rapidly. This was found to be the case for boson stars, gravastars, and superspinars, which are hypothetical stellar objects with an angular momentum larger than the Kerr bound~\cite{Pani:2009fd,Pani:2010jz}. Thus, the observation of GW events with large-spin mergers would favour a BH interpretation and put constraints on specific ECO models.

\subsubsection{Ringdown and quasi-normal modes}

During the last stage of the merger the compact object in the final state 
settles down to a stable form. This is the ringdown phase. The ringdown phase is interesting for all compact object collisions, and has been studied in detail for boson stars \cite{Macedo:2013qea,Macedo:2013jja,Macedo:2016wgh}, however as boson star signatures are likely to arise already in the modified inspiral and merger waveform, in this section we will instead focus on gravastars ($g^*$) as an example of ECOs that lead to distinctive predictions in the ringdown phase, different from those of BH. 

Before entering in the details of our discussion, let us mention the following important caveats. 
Apart from theoretical arguments, the reader should keep in mind that the measurement of a rapid event 
like the ringdown phase is a challenging experimental task.
In addition, as stressed in~\cite{Cardoso:2016rao,Barausse:2014tra}, only precision observations of the late-time ringdown signal can be used to 
unambiguously test the properties (or the absence) of an event horizon, especially for ECOs 
with compactness close to that of BHs.  The reason for this is that just after the merger the signal is dominated by oscillations of the light ring, which is the same for a BH and an ultra-compact object such as a $g^*$.  The light ring contribution does diminish, leaving the true quasi-normal modes in the later stages of the ringdown.  The importance of this effect 
depends on the compactness.  For objects with compactness very close to a BH, the light ring oscillations could give a ringdown mimicking a BH for many oscillations.

As discussed in \sect{sec:gravastar}, the simplest model of gravastars is described by three input parameters which, for convenience, we choose to be $\mathcal{P} \equiv \{ M,C,\delta\}$: the total gravitational mass $M$, the compactness $C\equiv M/R$, and the radius fraction in the outer shell $\delta \equiv (R-r_c)/R$.

\begin{figure}[t]
\begin{center}
\includegraphics[width=0.5\textwidth]{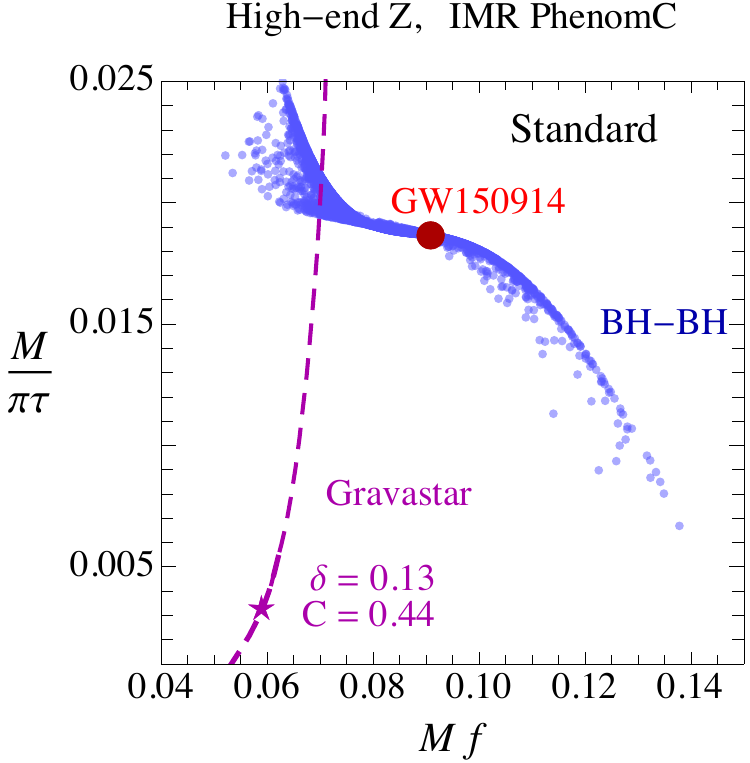}
\caption{\em Quasi-normal mode frequency and decay time for BH ringdown (blue dots) and $g^{*}$ ringdown (purple line).  Their clear separation shows that  the $g^{*}$ could be distinguished from the BH, if the ringdown phase is well measured.
The purple star refers to the special value of  quasi-normal mode frequencies used in \fig{fig:gStarWaveform}.
\label{fig:GravaStar}}
\end{center}
\end{figure}

Varying the parameters $\mathcal{P}$, we study the stability of the $g^*$ by solving the Oppenheimer-Volkoff equation describing -- in the context of general relativity -- the 
structure of a spherically symmetric body in static gravitational equilibrium.
We follow~\cite{Chirenti:2007mk}, where 
the conditions for non-rotating $g^*$ stability were discussed (see~\cite{Chirenti:2008pf} for a related discussion in the case of rotating $g^*$). We find that it is always possible to build a $g^*$ for any value of $\delta$ and any mass, as long as $C$ is smaller than about 0.44. The approach to the BH value of $C=0.5$ is possible, but requires small values of $\delta$. 

If GWs from a $g^*$ merger were observed, could the signal be distinguished from BHs? The crucial observation~\cite{Chirenti:2007mk,Chirenti:2016hzd} is that 
for a $g^*$ of mass $M$ the quasi-normal modes differ from those of a BH with the same mass (see also~\cite{Pani:2009ss,Pani:2010em}), and 
the origin of the difference between $g^*$ and BH quasi-normal modes is ultimately related 
 to the existence of an event horizon. We refer to ~\cite{Berti:2009kk} for a detailed discussion.

In the initial stages of the ringdown the signal will be dominated by oscillations of the light ring (see \cite{Cardoso:2016rao,Barausse:2014tra} for details), however in the later stages of the ringdown the waveform has the general structure
\begin{equation}\label{eq:QNM}
h_{\rm RD}(t) = \mathcal{A}_{\rm RD}(t)\cos\psi_{\rm RD}(t) = \sum_{n,l,m}\mathcal{A}_{\rm RD}^{(n,l,m)}(t)\cos\psi_{\rm RD}^{(n,l,m)}(t)~,
\end{equation}
in the time domain, where both amplitude and phase are decomposed in quasi-normal modes.
Quasi-normal modes describe oscillations that decay in time, 
and they represent the energy dissipation of a perturbed object.
\Eq{eq:QNM} can be recast in the form
\begin{equation}\label{eq:QNM2}
h_{\rm RD}(t) = \sum_{n,l,m}\theta(t)\mathcal{A}_{0}^{(n,l,m)}e^{-t/\tau_{nlm}}
\cos(2\pi f_{nlm}t)~,
\end{equation}
where $\tau_{nlm}$ is the decay constant of the mode and $f_{nlm}$ its quasi-normal frequency.\footnote{Note that in the frequency domain the
Fourier transform of the waveform in \eq{eq:QNM} is a Lorentzian function~\cite{Ajith:2009bn} 
\begin{equation}
\mathcal{L}(f,f_{\rm ring},\sigma) = \frac{\sigma/2\pi}{(f - f_{\rm ring})^2 + \sigma^2/4}~.
\end{equation}
 In the notation of~\cite{Ajith:2009bn} we have $\sigma = 1/(\pi \tau_{nlm})$, $f_{\rm ring} = f_{nlm}$.
}

For a Kerr BH the ringdown gravitational waveform is dominated by the quasi-normal mode with $n=0$, $l=m=2$ which is the most slowly damped 
mode.
For the merger of two Kerr BHs with total mass $M = M_1 + M_2$ and dimensionless rotation parameter $a_i\equiv J_i/M_{i}^2$ (where $J_i$ is the spin angular momentum), we find (omitting the indices $_{nlm}$)~\cite{Ajith:2009bn}
\begin{eqnarray}
\frac{M}{\tau} &=& \frac{\left[1-0.63(1-\chi)^{0.3}\right]}{4}(1-\chi)^{0.45} + \sum_{i = 1}^{3}\sum_{j = 0}^{N}y_{\tau}^{(i,j)} \eta^{i}\chi^{j}~,\label{eq:Freq1a}\\
2\pi fM &=& \left[1-0.63(1-\chi)^{0.3}\right] + \sum_{i = 1}^{3}\sum_{j = 0}^{N}y_{f}^{(i,j)} \eta^{i}\chi^{j}~,\label{eq:Freq2a}
\end{eqnarray}
where $\eta$ is the symmetric mass ratio given in \eq{eq:SymRatio} and the spins enter via the combination
\begin{equation}
\chi \equiv \frac{M_1a_1+M_2a_2}{M} ~.
\end{equation}
In eqs.~(\ref{eq:Freq1a})--(\ref{eq:Freq2a}) the first term represents the leading contribution in the
post-Newtonian expansion~\cite{Buonanno:1998gg}. In the post-Newtonian corrections the second sum is extended up to $N\equiv {\rm min}(3-i, 2)$, 
and the numerical coefficients $y_{\tau, f}^{(i,j)}$ are tabulated in ~\cite{Ajith:2009bn}.

For a $g^*$ the (axial) quasi-normal modes were computed in~\cite{Chirenti:2007mk}, and we use their results for our illustrative purposes.  In \fig{fig:GravaStar} we show the range of quasi-normal mode frequencies for BHs and for $g^*$. 
In order to populate
 the BH distribution, we computed the quasi-normal frequency and decay time in eqs.~(\ref{eq:Freq1a})--(\ref{eq:Freq2a}) using a simulated population of BH mergers from which we extracted, event by event, the masses $M_1$ and $M_2$.
  As far as the information about the spin is concerned, for each simulated event we randomly generated the two spin parameter $a_i \in [0,1]$.
  The label in \fig{fig:GravaStar} refers to the astrophysical assumptions underlying the specific BH population used in the plot (see \tab{tab:BinaryModels}). We refer to \sect{sec:census} for a detailed description of the numerical simulations of merging binaries used in this section and in the rest of the work.
 At this stage, the relevant message is that, since the BH and $g^*$ predictions mostly occupy different parameter regions, it appears that the two cases can be in principle distinguished.  For practical purposes we also consider whether these differences could be observed through the observed waveform.

\begin{figure}[t]
\begin{center}
\includegraphics[width=0.6\textwidth]{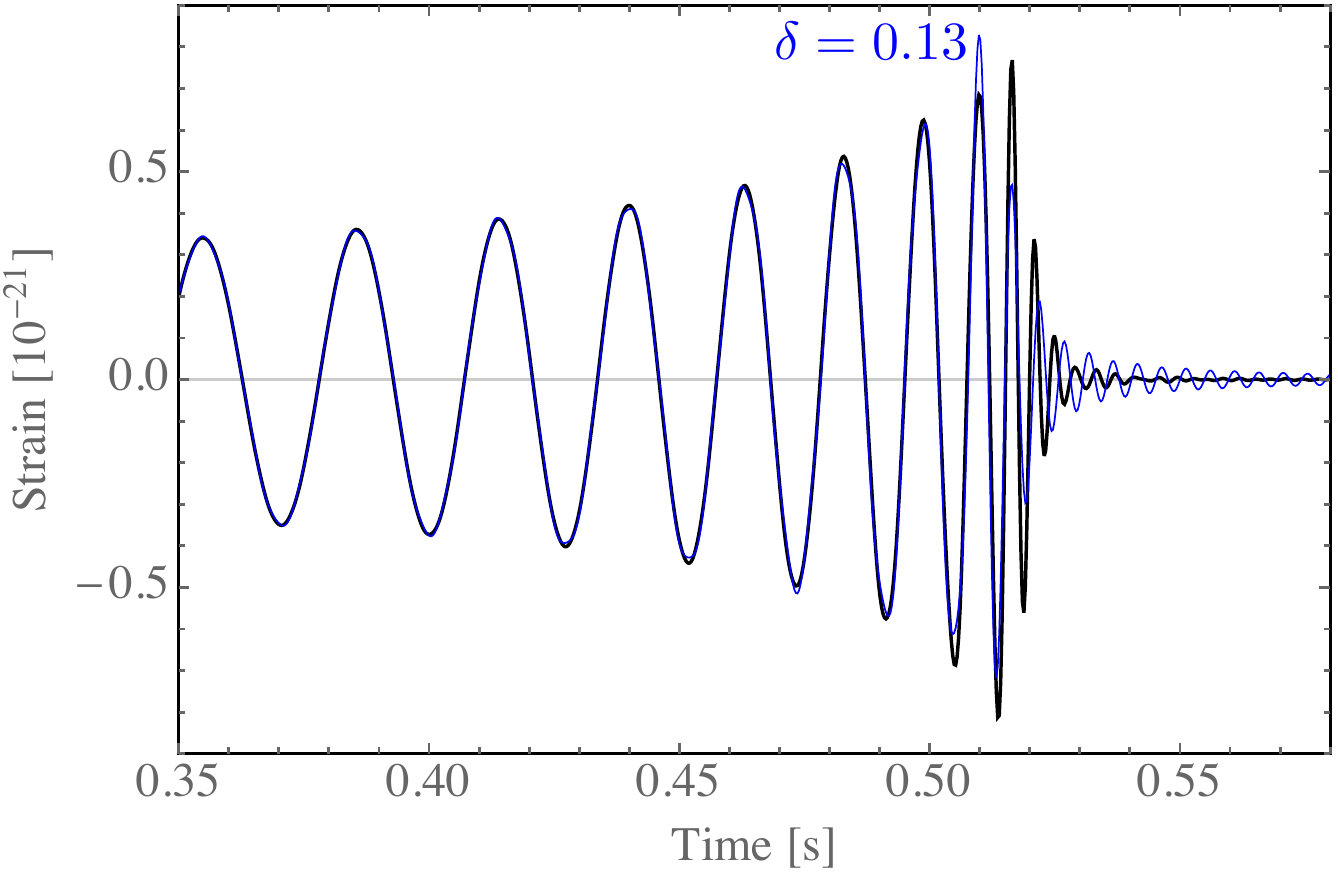}
\caption{\em Comparison between the waveform for a BH-BH merger with equal binary masses ($M_1=M_2=34.8\, M_\odot$, black), and a $g^*$ waveform with $C = 0.44$ and $\delta = 0.13$.  The ringdown parameters relative to the BH case were taken from \cite{Chirenti:2007mk}.
\label{fig:gStarWaveform}}
\end{center}
\end{figure}

In \fig{fig:gStarWaveform} we compare the waveform for a BH-BH merger with an estimated waveform for a $g^*$-$g^*$ merger, focussing on a $g^*$ with compactness $C = 0.44$ and thickness $\delta = 0.13$.  The $g^*$ waveform was obtained by taking the analytic waveform for a BH-BH merger from \cite{Ajith:2009bn} and replacing the BH ringdown frequencies with the $g^*$ ringdown frequencies.  While this comparison between the two may not be exact, it illustrates the difference in the ringdown component of the GW signature. It also suggests that precise measurements of the ringdown may lead to evidence for an ECO such as a $g^*$, and the signature would indicate an horizon-less object with  compactness similar to a BH.

 For the benchmark of \fig{fig:gStarWaveform}, where the compactness is $C=0.44$, the light-ring effects are likely to be important only in the first oscillation or so after merging, thus the quasi-normal modes in the ringdown would dominate and allow for discrimination from a BH as illustrated.\footnote{We are particularly grateful to Vitor Cardoso and Paolo Pani for enlightening discussions on GW ringdown signatures.}  However, in general, while an exotic ringdown may give evidence for an exotic final state, on the converse the lack of an exotic ringdown does not necessarily imply that the object was not exotic.
Only future observations will shed light on the real constraining power of ringdown measurements.

\subsection{Features in a Binary Mass Census}
\label{sec:census}

In \sect{sec:waveforms} we focused on ECO signatures that can be extracted from the waveform of a single GW event. Here we discuss the physics information that can be extracted from statistical distributions of GW events, which will be available once a significant number of mergers have been observed.

We also remark that, even if ECO signatures were identified in single waveforms, distributions of events will be helpful to understand the ECO mass function and possibly gain knowledge about their formation.  To explore this possibility we must first understand the expected distribution of known compact objects, as this will form the known `background' distribution.

\subsubsection{On the mass distribution of compact merging binaries in the local Universe}

The mass distribution of compact object binaries can be effectively organised in a plane of the chirp mass $M_c$ and the symmetric mass ratio $\eta$.  While the chirp mass is defined in \eq{eq:chirp}, the symmetric mass ratio is related to the primary and secondary binary constituent masses $M_1,M_2$ as in \eq{eq:SymRatio}, and it 
 corresponds to the ratio between reduced and total mass.

\begin{table}[!htb!]
\begin{center}
\begin{tabular}{||c||c||}
\hline\hline
\textbf{Model} & \textbf{Main physical properties and parameter variations}\\  
\hline\hline
\multirow{3}{*}{Standard} & $M_{\rm NS}^{\rm max} = 2.5$ $M_{\odot}$, {\it rapid} SN explosion, $\lambda =$ {\it Nanjing}~\cite{Xu:2010wx}, \\
    &  standard NS kick  $\sigma = 265$ km s$^{-1}$, reduced BH natal kick, \\
    & HG CE donor not allowed, high-end metallicity  \\
\hline
 Optimistic CE & HG CE donor allowed   \\
\hline
Delayed SN & Delayed model for supernova explosions   \\
\hline
High BH kicks & Full natal kick for BH \\
\hline\hline
\end{tabular}
\end{center}
\caption{{\it  Models of stellar binary evolution presented in~\cite{Dominik:2012kk,Dominik:2013tma,Dominik:2014yma}.
Each variation is defined by changing one parameter, which we specify  in the second column, w.r.t. the `Standard' model.  Numerical simulations are available at~\cite{syntheticuniverse}.}}

\label{tab:BinaryModels}
\end{table}

\begin{figure}[!htb!]
\minipage{0.5\textwidth}
  \includegraphics[width=.9\linewidth]{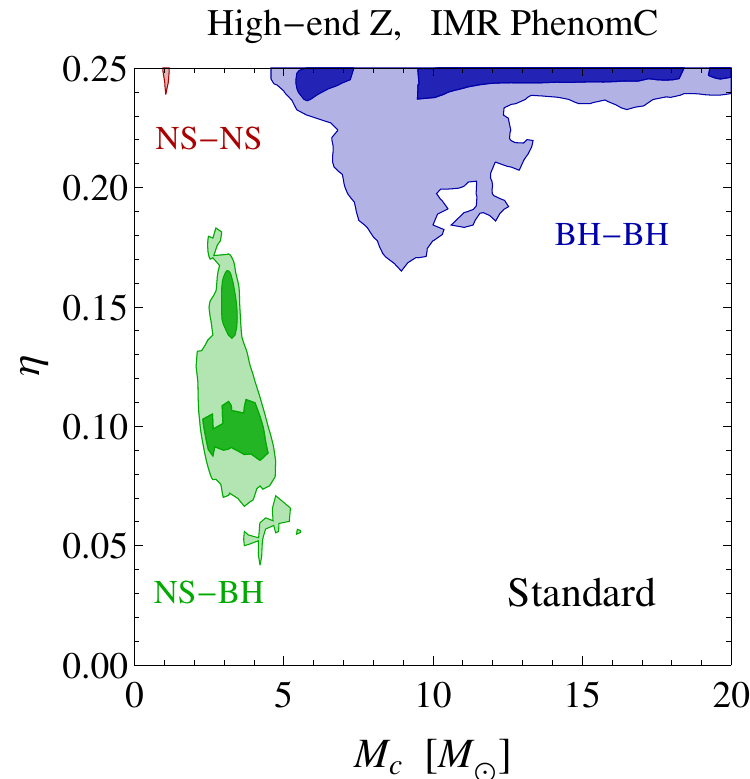}
\endminipage\hfill
\minipage{0.5\textwidth}
  \includegraphics[width=.9\linewidth]{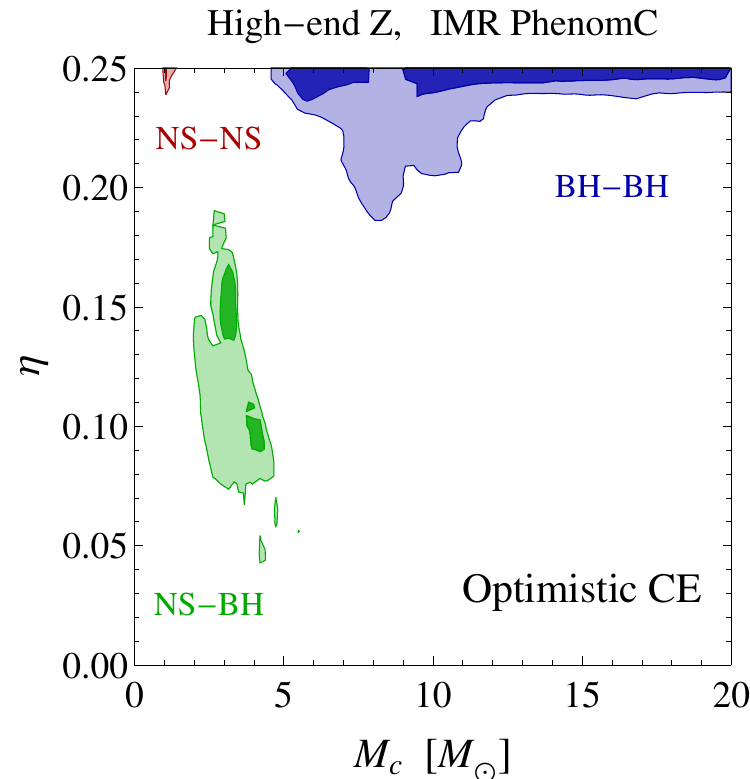}
\endminipage \vspace{.25 cm}\\
\minipage{0.5\textwidth}
  \includegraphics[width=.9\linewidth]{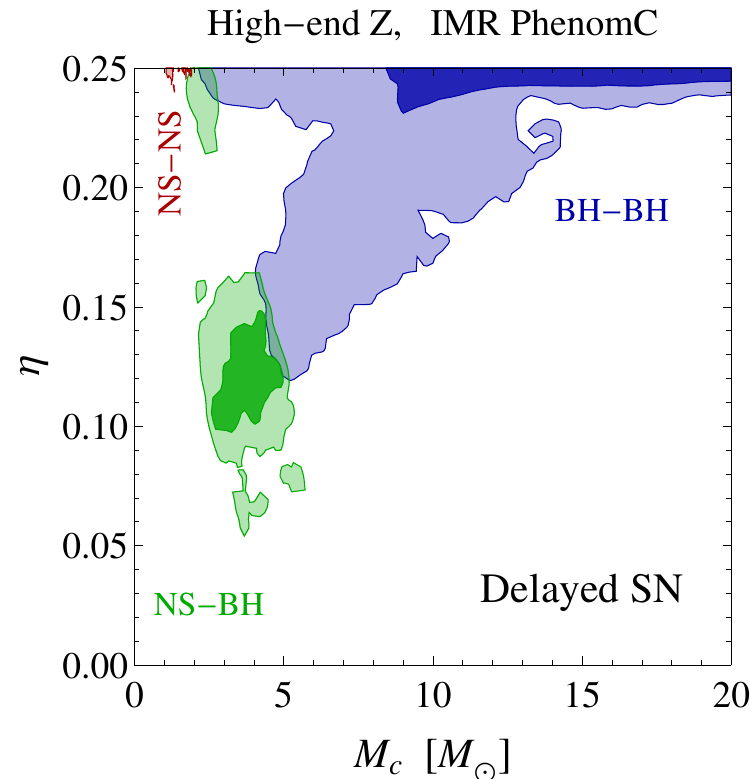}
\endminipage\hfill
\minipage{0.5\textwidth}
  \includegraphics[width=.9\linewidth]{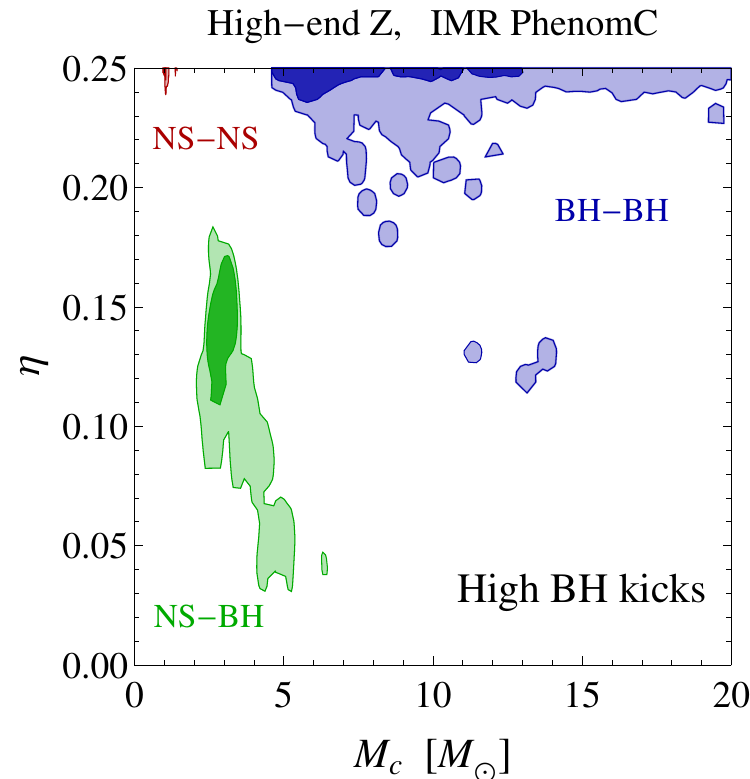}
\endminipage 
\caption{\em 
Binary mass distribution -- in the plane chirp mass vs. symmetric mass ratio $(M_c, \eta)$ -- for the four stellar evolution models considered in \tab{tab:BinaryModels}.
For each distribution the contours of different colour shading represent the $68\%$ and $99\%$ CL regions.
}
\label{fig:TestMass}
\end{figure}

When the population of mergers is projected onto the $M_c-\eta$ plane, binaries involving neutron stars and black holes (in the combinations NS-NS, NS-BH, and BH-BH) generally populate different regions \cite{Mandel:2015spa}.  
The mass distribution of these `conventional'  compact object binaries depends on the assumed model of stellar binary evolution.
In turn, models of  stellar binary evolution depend on a number of parameters that are largely unconstrained or poorly known.
In the following we briefly mention some of the most relevant sources of uncertainty and features of these models, while we 
refer the reader to~\cite{Dominik:2012kk,Dominik:2013tma,Dominik:2014yma}  for a more detailed discussion.

\emph{Details of the `common envelope (CE) phase'.}  The formation of a compact binary object is preceded by an evolutionary phase in which the binary system 
evolves enclosed into a gas envelope originated from the interactions between the donor and the companion star.
The most important physical quantity  describing the CE phase is the binding energy of the envelope, which in turn depends on the characteristic  mass and radius of the binary system.
If the binding energy is too high  the stars will merge during the CE phase thus preventing the formation of a compact binary object.
If the binding energy is too low
the decrease in orbital separation during the CE phase  is too slow, 
and the 
resulting compact binary object  will never merge  within an Hubble time. 
Furthermore, the formation of a CE strongly depends on the properties of the donor star. 
This is particularly true if the donor star is situated in the Hertzsprung gap (HG) of the Hertzsprung-Russell diagram; in this case,  two alternative choices are commonly adopted:
either HG stars are CE donors or they are not.

\emph{Unknown maximum mass of NS and minimal mass of BH.} 
As discussed in \sect{sec:conv}, all observed NS have masses below 2 $M_{\odot}$, and theoretical modelling suggests that this is likely to be an upper bound. On the other hand, stellar BH, which are expected to be generated by the gravitational collapse of massive stars, have a mass distribution known to  start from about 5 $M_{\odot}$ and to extend to tens of solar masses. In other words, observations (mostly based on X-ray measurements) favour the existence of a {\it mass gap} between the distribution of NS and stellar BH~\cite{Bailyn:1997xt,Fryer:1999ht,Ozel:2010su,Farr:2010tu}, although it cannot be excluded that this is only the result of some observational bias. 

The presence of this mass gap may seem surprising, given that the progenitor stellar mass distribution is known to be smooth. Hence, the distribution of the ensuing compact remnants might be expected to be smooth as well. However, 
in ref.~\cite{Belczynski:2011bn} it was shown that the presence of this mass gap can be linked to the evolutionary mechanism of dying stars. For sufficiently fast growth of the instabilities causing the stellar death, the evolution leads to two distinct outcomes: either a violent explosion that ejects most of the mass from the star, leaving a NS remnant; or a failed supernova, in which all the stellar mass collapses in a BH. A mass gap is dynamically explained, as long as the growth timescale is sufficiently short (less than about 200 ms).

\emph{Physics of the explosions that form compact objects.} The physics describing   
star evolution, supernova explosions and the subsequent formation of compact objects (either NS or BH)
is plagued  by many uncertainties.
{\it i)} The evolution of a main sequence star toward its eventual supernova stage is affected by mass loss via stellar winds.
{\it ii)} Supernova explosions may be modelled by means of different mechanism, depending on the type of instability 
that eventually lead to the ejection of in-falling matter.
{\it iii)} Finally, the formation of a NS or BH from supernova explosions may be accompanied by a natal kick.

\emph{Metallicity $Z$.} The information about the metallicity -- which in general describes 
the fraction of mass that is not in Hydrogen or Helium -- is of fundamental importance to model the chemical evolution of the local Universe.

The impact of these uncertainties on the expected distribution of binary merger primaries may be estimated using simulations.  In \tab{tab:BinaryModels}, we summarise the features of different models, including the `Standard' model and three variations.  The resulting distributions in the $M_c - \eta$ plane for these models are shown in \fig{fig:TestMass}.  Clearly, variations in the underlying model do lead to differences in the expected distributions. However, a qualitative feature that persists across all of these scenarios is that NS-NS, NS-BH, and BH-BH mergers populate reasonably distinct regions.  It is then tempting to conclude that, if
future GW observations fell outside the regions in \fig{fig:TestMass} covered by `conventional' mergers, we could infer evidence for an exotic population of compact objects.  However, to assess this possibility it is necessary to consider more carefully the GW observables.

\begin{figure}[!t!]
\minipage{0.5\textwidth}
  \includegraphics[width=.9\linewidth]{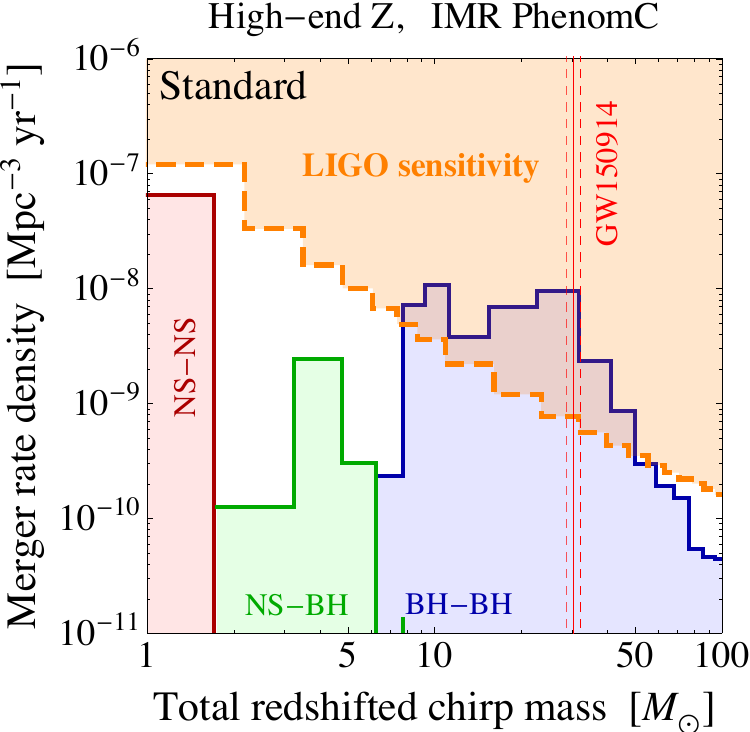}
\endminipage\hfill
\minipage{0.5\textwidth}
  \includegraphics[width=.9\linewidth]{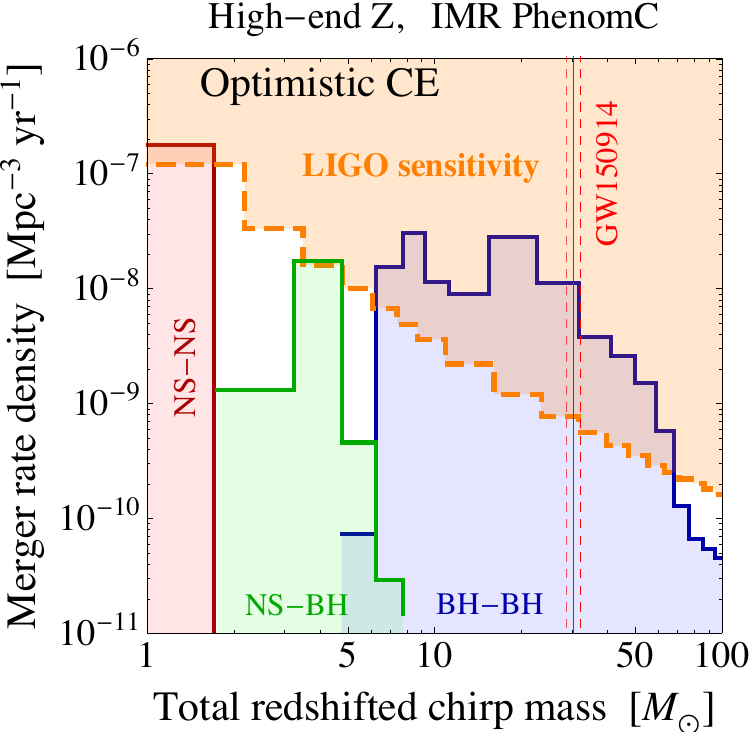}
\endminipage \vspace{.25 cm}\\
\minipage{0.5\textwidth}
  \includegraphics[width=.9\linewidth]{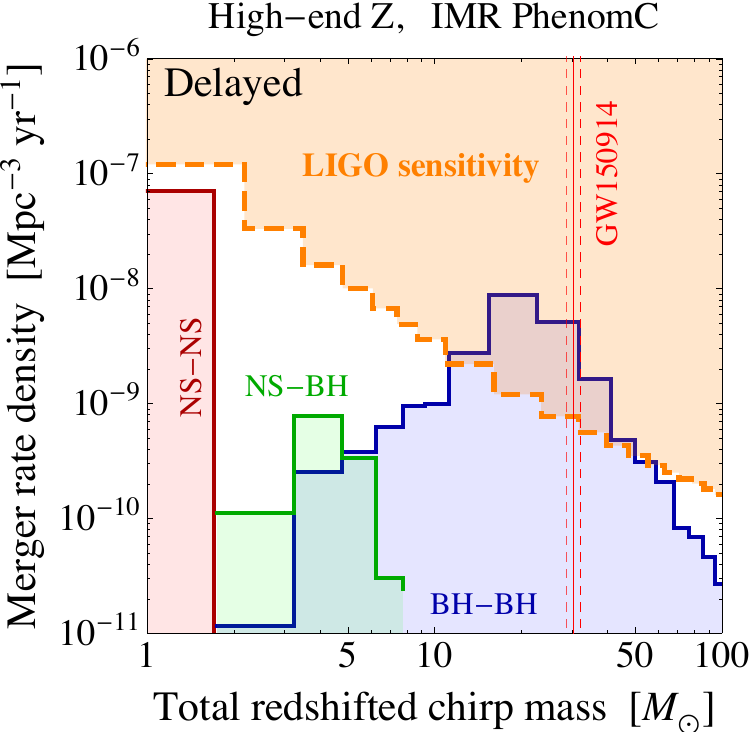}
\endminipage\hfill
\minipage{0.5\textwidth}
  \includegraphics[width=.9\linewidth]{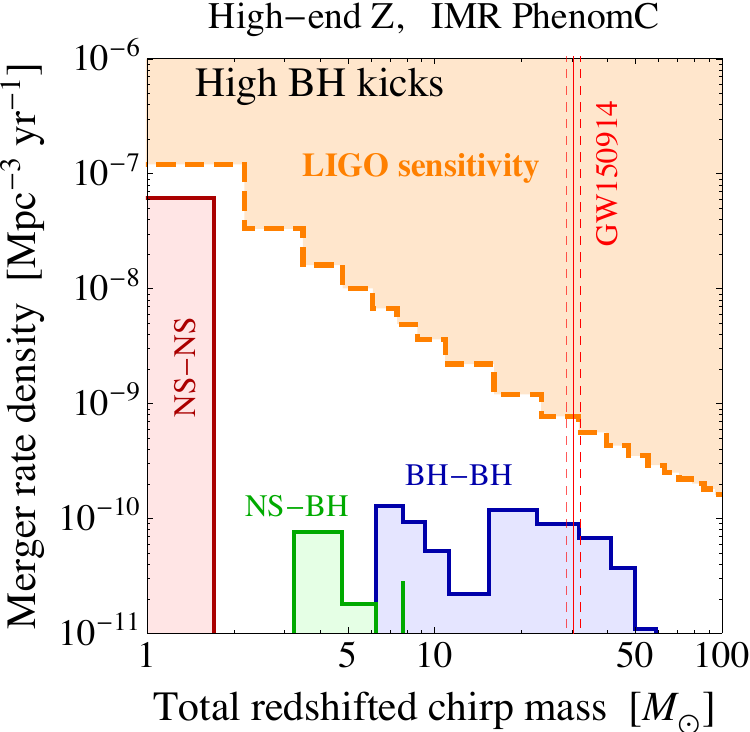}
\endminipage 
\caption{\em 
Merger rate density as a function of the redshifted chirp mass for the four stellar evolution models presented in \tab{tab:BinaryModels}.
In each case, we show the distributions for NS-NS, NS-BH and BH-BH event rates. 
The vertical red line corresponds to the event GW$150914$, with measured detector-frame chirp mass $M_c = 30.5^{+1.7}_{-1.8}$ $M_{\odot}$
and source redshift $z= 0.093^{+0.028}_{-0.036}$~\cite{TheLIGOScientific:2016wfe}. Also shown is the LIGO detection sensitivity, obtained with the estimate described in the text.
}
\label{fig:Test}
\end{figure}

A GW observatory will observe the GW emitted in a binary merger.  Physical properties of the binary system are then extracted by comparing the observed waveform with waveforms predicted within general relativity.  Thus $M_c$ and $\eta$ are not themselves observables, but they may be determined from this comparison.  $M_c$ is relatively universal and may be extracted from the inspiral phase of the wave.  However, an independent extraction of $\eta$ requires consideration of the merger phase and is thus only accessible by comparison with waveforms generated numerically.  If a detected waveform is studied by comparing with expected NS-NS, NS-BH, and  BH-BH waveforms, then only best-fit parameters populating these regions in the $M_c-\eta$ plane will be found.  If the waveform turns out sufficiently exotic that none of these waveforms give a good fit, then discovery would come from single-event analysis, as described in \sect{sec:waveforms}.  Thus, it seems unlikely that evidence for ECOs would first arise by analyses of distributions in the $M_c-\eta$ plane. 

However, the chirp mass may be extracted from the inspiral phase and thus does not require a detailed comparison between the observed wave and the numerical GR waveforms generated for specific scenarios.  Thus, once many merger events have been observed, it may be possible to search for a new distribution of ECOs by considering the chirp mass alone, without relying too heavily on specific expected waveforms.  In \fig{fig:Test} we show the four distributions of  \fig{fig:TestMass} projected onto the chirp mass axes alone.
To make a concrete connection with observations, we superimpose the expected LIGO sensitivity as a function of the chirp mass.
This sensitivity was computed in a way similar to~\cite{Belczynski:2015tba}.  At each value of red-shifted chirp mass --  under the simplifying assumption $M_1 = M_2$ -- we computed the maximum luminosity distance at which the signal-to-noise ratio in \eq{intrho} reaches the detectability 
threshold $\rho = 8$. The corresponding redshift, evaluated according to \eq{eq:RedShift}, allows us to reconstruct the maximum volume $V$ inside which the analysed 
binary could be seen. The expected upper limit on the merger rate density, after an operating time $T$, is $\mathcal{R} = 1/(\epsilon VT)$, where $\epsilon$ is an experimental efficiency which we take to be 30\%. 

Note that this result, although strictly valid only for the case of equal-mass BH-BH mergers, should give a correct order-of-magnitude estimate of the expected LIGO sensitivity.

   Clearly, there are large variations in the expected distributions based on the underlying model.  However, in three of the four choices an identifiable chirp mass gap appears between the NS-NS and BH-BH distributions.  This mass gap does not arise in the model with delayed SN explosion, as also clear from 
the corresponding mass distribution in \fig{fig:TestMass}. However, when projected onto the chirp mass axes, 
the merger rate density of BH-BH binaries populating the mass gap turns out to be extremely low, order of magnitudes below the LIGO sensitivity.
Finally, we note that the BH-BH binary merger rate expected from the scenario with high natal BH kicks is smaller than for the other models and appears to be in some tension with the observation of GW150914, given the LIGO sensitivity.  Thus, the high natal BH kick scenario already appears disfavoured, as also concluded in \cite{Belczynski:2016obo}.  These simulations thus seem to suggest that a mass gap is likely for binary mergers of known objects.

There is no reason to expect ECOs to have a mass distribution that exhibits a mass gap in precisely the same location. Thus, events filling the gap in the chirp mass distribution would point towards a population of ECOs.  This is especially true if the waveforms appear to be of the NS-NS or NS-BH type, because massive ECOs with compactness $C<1/2$ could resemble NS, although with anomalously large masses.

To make a quantifiable assessment of this possibility would require not only a better understanding of the systematic uncertainties in the expected distribution of known objects, but also concrete predictions for the distribution of ECOs.  As this distribution will depend in detail on the formation mechanism it is highly model dependent and would require treatment on a case-by-case basis for each particle physics model.  As this is beyond the scope of this work we will simply speculate that the comparison of the observed distribution of chirp masses with that expected from theoretical models can give some information about a possible population of ECOs. Unfortunately, the large astrophysical uncertainties in the modelling of the `background' are a severe limiting factor for the discovery of a `signal'.

\section{Test of the Area Theorem}
\label{sec:area}
Hawking's area theorem~\cite{Hawking:1971tu} states that the sum of the horizon areas of a system of BHs never decreases. This remarkable result follows directly from general relativity and the null energy condition. However, once we use the thermodynamical identification of a BH temperature (inversely proportional to mass, $T=1/8\pi M$~\cite{Hawking:1974rv,Hawking:1974sw}) and BH entropy (proportional to area, $S=A/4$~\cite{Bekenstein:1972tm,Bekenstein:1973ur,Bekenstein:1974ax}), Hawking's theorem acquires a new interpretation~\cite{Jacobson:1993vj,Bekenstein:1994bc,Jacobson:1995uq}. The statement
about the ever increasing area can be viewed as a generalised second law of thermodynamics for the ever increasing entropy. 

This thermodynamical interpretation of the area theorem has a deep physical significance that goes beyond classical general relativity. Indeed, the area theorem is apparently violated by the quantum process of Hawking's radiation. During BH evaporation, conservation of energy implies that the BH mass must decrease, hence the BH area must decrease as well. This contradiction with the area theorem is resolved in the context of the generalised second law of thermodynamics. Once we account for the entropy of the emitted matter and radiation, and not only the entropy associated with the BH area, the second law of thermodynamics is respected by BH evaporation. All this to say that any observational test of the area theorem is a way of probing the fundamental principles of general relativity and statistical mechanics.

It is remarkable that GW detection can indeed provide a test of the area theorem. The horizon area of a Kerr BH 
of mass $M$ and angular momentum $J$
is given by
\beq
A=8\pi M^2 \left( 1+\sqrt{1-a^2}\right) ~~~~~a\equiv \frac{J}{M^2}\, .
\eeq
For the process of two BH with horizon areas $A_{1,2}$ merging into a final BH of area $A_f$, Hawking's theorem predicts
\beq
A_f > A_1+A_2 ~~~~~~{\rm (area~theorem)}\, .
\label{ineq}
\eeq

The area theorem does not only provide a test of fundamental principles, but can also probe the existence of ECOs.
In order to understand how ECOs could give an apparent violation of the area theorem, if erroneously taken for BHs, it is useful to clarify the physical content of the area theorem when applied to merger events. 
\eq{ineq} can be written as a lower bound on the final BH mass
\beq
M_f > \sqrt{M_1^2\, s_1 + M_2^2\, s_2}
 \, , ~~~~~~~s_{1,2}\equiv \frac{1+\sqrt{1-a_{1,2}^2}}{1+\sqrt{1-a_{f}^2}} \, .
 \label{spotorno}
 \eeq
 The absence of naked singularities (aka cosmic censorship conjecture~\cite{Penrose:1969pc}) requires $0\le a \le 1$, which implies $s_{1,2}>1/2$.
Using this condition in \eq{spotorno}, we find
\beq
M_f >  \sqrt{\frac{M_1^2+M_2^2}{2}}\, .
\eeq
In words, {\it Hawking's theorem implies a simple lower bound on the mass $M_f$ of the BH resulting from the merging of two BHs with masses $M_{1,2}$}.

Another way of reading \eq{ineq} is as an inequality on the efficiency in GW emission, $\epsilon \equiv (M_1+M_2-M_f)/(M_1+M_2)$,
\beq
\epsilon < 1-\frac{\sqrt{s_1+r^2\, s_2}}{1+r} \, , ~~~~~~~r\equiv \frac{M_2}{M_1}
 \label{efficiency}
 \eeq
The upper bound on $\epsilon$ is maximised for maximally-rotating initial BHs and non-rotating final BH ($s_{1,2} =1/2$), giving $\epsilon < 1-\sqrt{(1+r^2)/2}/(1+r)$. In turn, this is optimised for $r=1$, giving $\epsilon <1/2$. In the special case of slowly-rotating initial BHs ($1\le s_1=s_2 \le 2$), the maximum upper bound is  $\epsilon < 1-\sqrt{1+r^2}/(1+r)$, which is optimised for $r=1$ giving $\epsilon <1-1/\sqrt{2}$.

To summarise, {\it the area theorem prescribes a maximum efficiency in GW emission for the merging process of two BHs}. An apparent violation of the theorem could occur if the merging objects are not really BHs and emit other forms of energy, which goes undetected and exceeds the bound in \eq{efficiency}. In other words, the apparent violation of the area theorem would occur in the same way as for Hawking's BH evaporation. Emission of non-gravitational radiation leads to entropy production, left unaccounted for by the area theorem. As a result, the area theorem is not satisfied, although
the generalised second law of thermodynamics holds valid. 

As a concrete example of observational test of the area theorem, let us consider the detection of GW150914, for which the extracted parameters are $M_1 =36^{+5}_{-4}\, M_\odot$ and $a_1=0.3^{+0.5}_{-0.3}$ for the primary and $M_2 =29\pm 4\, M_\odot$ and $a_2=0.5^{+0.5}_{-0.4}$ for the secondary BH~\cite{TheLIGOScientific:2016wfe}.
As far as the final state BH is concerned, we use the 90\% credible regions  for the mass $M_f$
and dimensionless spin $a_f$ provided in~\cite{TheLIGOScientific:2016src}. 
This gives
\beq
A_1 + A_2 =(2.2\pm 0.2) \times 10^5\ {\rm km}^2 \, , ~~~~~~
A_f =(3.8\pm 1.3) \times 10^5\ {\rm km}^2 \, ,
\label{error}
\eeq
where we have taken into account the error correlation among the input parameters using the information in~\cite{TheLIGOScientific:2016wfe,TheLIGOScientific:2016src}. 

\begin{figure}[!htb!]
\centering
  \includegraphics[width=.5\linewidth]{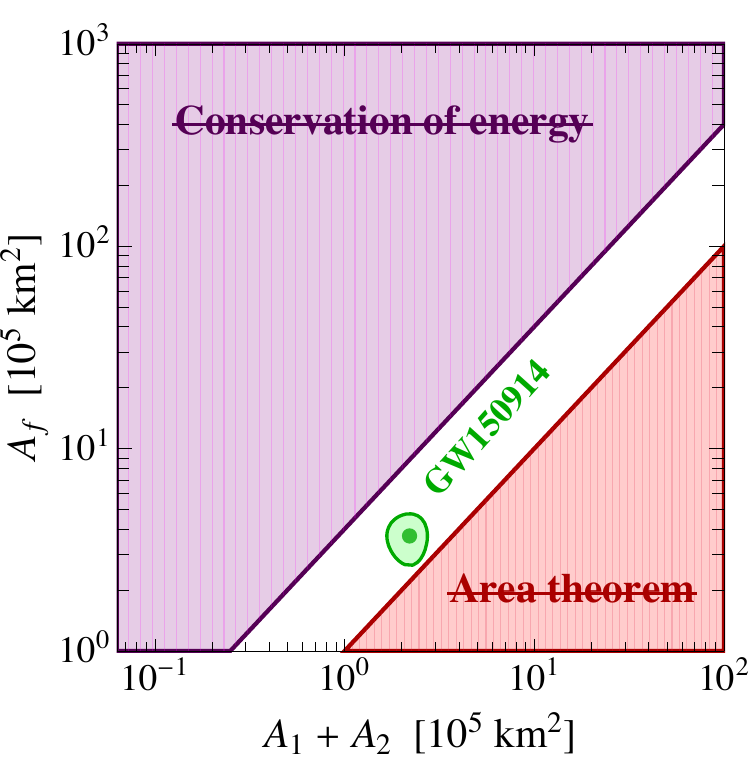}
\caption{\em In the plane of the horizon areas of initial ($A_1+A_2$) and final ($A_f$) BH binary mergers, the dashed regions indicate where Hawking's area theorem is violated $[A_f<A_1+A_2]$ and where energy conservation does not hold $[A_f> 4(A_1+A_2)]$. The result from GW150914 corresponds to the green area (at 90\% CL) and lies within the allowed region (white band). The correlation between the extraction of $A_f$ and $A_1+A_2$ is not taken into account.
}
\label{fig:AreaTheorem}
\end{figure}

The result in \eq{error} shows that the observation of GW150914 has verified \eq{ineq}, hence Hawking's 
area theorem, well within errors. This is also visualised in \fig{fig:AreaTheorem}, where we show how the LIGO result falls within the region of plane $A_1+A_2$ vs $A_f$ allowed by the area theorem. In the same plane we also show the region excluded by the condition on conservation of energy $M_f<M_1+M_2 $, which corresponds to $A_f < ( \sqrt {A_1/s_1} + \sqrt {A_2/s_2} )^2$. Recalling that $s_{1,2}>1/2$, we obtain
\beq
A_f<4\, (A_1+A_2)~~~~~~{\rm (energy~conservation)} \, .
\label{ineqarea}
\eeq

Masses and spins of both initial and final state BHs were obtained in~\cite{TheLIGOScientific:2016wfe} by comparing (via a likelihood fit)
 the observed signal with 
waveform templates generated in the context  of numerical relativity through a full simulation of the inspiral, merger and ringdown phases.
It should be stressed that this experimental determination of $A_f$ is not independent of $A_1+A_2$. Once one assumes that the merging binaries are BH and fixes the initial parameters ($M_{1,2}$, $a_{1,2}$, the relative angles between the spin vectors, the angular velocity, and orbit eccentricity), the values of $M_f$ and $a_f$ are uniquely determined by the general relativistic calculation. A genuine test of the area theorem would require an independent extraction of the initial and final BH parameters. This is not impossible as the inspiral and ringdown phases are mostly sensitive to initial and final BH parameters, respectively.   
For this reason,  for the final state BH we used the measurements of $M_f$ and $a_f$ obtained in~\cite{TheLIGOScientific:2016src} with only post-inspiral informations.
This choice allows us to introduce a certain degree of independence between initial and final state measurements thus providing a non-trivial test of the area theorem.

Future GW observations of BH mergers will further test the area theorem. What if these observations do not agree with the theorem's expectations? Unless one is ready to abandon the basic principles of general relativity, measurements of apparent violations of the area theorem can be used to argue for the existence of ECOs.
Depending on their nature, ECOs could emit significant fractions of their energy in non-gravitational forms, in practice exceeding the bound in \eq{efficiency}. This can happen if they emit intense electromagnetic radiation in collimated narrow beams not pointing towards the line of sight, or if they radiate particles (such as dark photons or light axion-like particles) that remain invisible to us.

The possibility of testing the area theorem has a theoretical interest independent of ECOs, since its validity is not guaranteed in various extension of general relativity (GR).
 The key observation for proving the theorem is that in GR the null energy condition -- that is $T_{ab}k^a k^b \geqslant 0$ for any null vectors $k^a$ -- implies, as a direct consequence of Einstein equations ($R_{ab} - g_{ab}R/2 = 8\pi G_N T_{ab}$), the null convergence condition $R_{ab}k^a k^b \geqslant 0$. In turn, the null convergence condition ensures, by means of the Raychaudhuri equation describing geodesics congruence, the validity of the area theorem. The proof is not valid in extended gravity 
theories, when the field equations differ from those of GR. This is the case of the Brans-Dicke theory (and, more generally, scalar-tensor gravity)~\cite{Kang:1996rj,Faraoni:2010yi}, in which Einstein equations are modified by new terms beyond the energy-momentum tensor involving the scalar field.

As a consequence, the null energy condition does not imply anymore the null convergence condition, thus invalidating Hawking's argument.
Even if static asymptotic BH must recover GR solutions (the no-hair theorem), 
during the dynamical stages of the merging process the validity of the area theorem is not guaranteed.
In terms of energy balance, this corresponds to the emission -- in addition to tensor GW perturbations -- of the extra scalar degrees of freedom.

In conclusion, the search for violations of the area theorem
can be used as a tool to confirm potential ECO discoveries and to derive information on undetected forms of radiation. Furthermore, testing the area theorem is an interesting task independently of searches for ECOs, as it offers a probe of fundamental principles.

\section{Summary}
\label{sec:conc}
The discovery of GWs from a BH-BH merger marked a breathtaking advance in our understanding of nature.  It signalled a new trajectory in astronomical and gravitational science.  To determine the role of such observations within particle physics it is necessary to connect the microscopic with the astronomical.

This work has made a step in this direction by determining scenarios in which well motivated new particles beyond the Standard Model may coalesce into solar-mass-scale exotic compact objects.  To this end, \sect{sec:exo} reviewed a number of new physics scenarios which predict dark particles able to give rise to ECOs with mass and compactness accessible to LIGO observations. While dark matter is our primary motivation, we also stress that light and stable particles, interacting only feebly with ordinary matter, are not unusual in many well motivated extensions of the Standard Model. It is conceivable that these relic particles -- or at least a fraction of them, whether explaining dark matter or not -- may clump in celestial bodies of astronomical size. Dark matter or exotic matter could then coexist in the form of microscopic dust, as unbound elementary particles, and dark macroscopic bodies, as ECOs. No experimental observation excludes this theoretical scenario. Limits from gravitational microlensing allow for ECO galactic populations as dense as ordinary stellar populations for ECO masses below about 10 $M_\odot$, while these limits quickly evaporate for heavier ECOs. 

With this broad overview in hand, in \sect{signatures} the observable signatures of new-physics motivated ECOs were considered. In \sect{sec:Frequency} we determined LIGO sensitivity to ECO merger detection. First, we have expressed our results in a model-independent way, presenting them in terms of the two astronomical ECO parameters most relevant for GW emission: the total ECO mass $M$ and its compactness $C$ (see \fig{fig:SNR}). Next, we have studied three representative examples of ECOs in specific microphysics setups: boson stars composed of a scalar field with repulsive self-interaction, boson stars composed of a free scalar field, and fermion stars composed of a spinor field with repulsive interaction mediated by a light vector mediator. In each case, we have identified the LIGO sensitivity range in terms of microscopic parameters, such as the dark particle mass and couplings (see \fig{fig:BosonStarScan} for the interacting boson, \fig{fig:FreeBosonStar} for the free boson, and \fig{fig:FermionStarScan} for the fermion). The intriguing result is that the values of the particle masses accessible to LIGO lie in the same ballpark as the mass predictions from asymmetric dark matter models (for interacting bosons or fermions) and span an interesting range for axion-like particles (for free bosons). These considerations prove that LIGO is suited to searching for well-motivated dark matter models, if dark matter has a dust component (in the form of free-roaming elementary particles) and a stellar component (in the form of clumped ECOs).     

In \sect{sec:waveforms} we study how evidence for ECOs could emerge from the detection of a single GW event, through the observation of distinctive features in the gravitational waveform. A key point is that the only known compact objects believed to have mass above about 5 $M_\odot$ are black holes.  BH-BH mergers have very characteristic waveforms related to the fact that they are both maximally compact and have an event horizon.  Mergers involving an ECO would exhibit exotic waveforms, typically distinguishable from BH-BH mergers.  Thus, if the waveform from a merger involving two very massive stellar objects were observed to be significantly different from the BH-BH case, this would unambiguously indicate an unexpected population of ECOs and herald the existence of new particles which may be connected with the dark matter puzzle. 

We discuss two examples of deviations from BH waveforms. First, we study the impact of the size of the ECO, considering a fermion star with compactness smaller than the Schwarzschild value. Although the beginning of the inspiral phase is indistinguishable from the case of BHs with the same masses as the ECOs, deviations appear at the onset of the merger phase when the objects are sufficiently close for size effects to become relevant (see \fig{fig:NSNS}). In the second example we consider gravastars, hypothetical astrophysical objects made of dark energy kept together by the gravitational pull of an outer massive shell. In this case, although the gravastar ECO is a good mimicker of BHs, differences in the waveform appear in the ringdown phase, when the newly-formed combined object settles down to a stable configuration (see \fig{fig:gStarWaveform}). The absence of an horizon in the gravastar ECO, as opposed to the BH, leads to distinctive features in the quasi-normal modes.

After having studied what can be learned from a single GW observation, in \sect{sec:census} we turn our attention towards features in the statistical distribution of GW events. Combining astronomical observations and theoretical modelling, the mass distribution of BHs and NSs can be predicted, although large uncertainties are still present. One emergent feature is a mass gap between the NS distribution, which extends only up to about 2 $M_\odot$, and the BH distributions, which appears to start at about 5 $M_\odot$. As reviewed in \sect{sec:census}, the existence of this mass gap, besides being so far verified by observations, may have some theoretical justifications. If a population of ECOs exists in the universe, there is no reason why it should not fill this gap. Moreover, an ECO population could be identified as an anomalous feature in the distribution of detected GW events, which could not be explained in terms of BH or NS expectations (see \fig{fig:Test}). Unfortunately, at present, lack of knowledge about the ECO formation processes prevents us from making definitive predictions on the ECO mass function.
However, even if waveform measurements seem a more promising tool for ECO discovery, studies of event distribution carry unique information, which may become essential to understand the origin of ECOs in the universe and their formation mechanisms. 
Moreover, an important conclusion can be drawn by the fact that any astrophysical body made of ordinary matter is expected to be a BH, if heavier than about 2 $M_\odot$. This means that any GW observation of objects heavier than 2 $M_\odot$ that do not match a BH template will give immediate evidence for an ECO discovery.

In \sect{sec:area} we show how GW detection from BH mergers can be used to test Hawking's area theorem~\cite{Hawking:1971tu}. Since the theorem relies on the basic assumptions of general relativity and the null energy condition, these tests can be viewed as a probe of fundamental principles. In the case of binary BH mergers, the area theorem translates into a forbidden region in the plane $A_1+A_2$ vs $A_f$, where $A_{1,2,f}$ are the horizon areas of the two initial BHs and final BH, respectively. Conservation of energy rules out another portion of this plane, leaving only a band, which remains theoretically acceptable (see \fig{fig:AreaTheorem}). Comfortingly, the GW150914 observation lies safely in the allowed band. It will be useful to see how future observations confirm this result. In our context, tests of the area theorem provide a useful tool to probe ECO properties. Indeed, taking for granted the validity of general relativity, apparent violations of the area theorem occur if the merging objects emit a large amount of undetected radiation. This is possible in certain ECO models in which the microscopic constituents can dissipate energy in the form of dark photons or other hidden particles. Tests of the area theorem can then be translated into searches for emission of dark radiation.

It is not yet firmly established how the new gravitational window into the hidden Universe provided by solar-mass GW astronomy may be exploited to further our understanding of physics beyond the Standard Model.  However, this work demonstrates a clear synergy between future GW observations and the search for new fundamental particles.

\subsubsection*{Acknowledgments}
We wish to thank Vitor Cardoso, Paolo Pani, and Antonio Riotto for useful comments.

\end{document}